\documentclass[twocolumn,aps,prd,preprintnumbers,nofootinbib]{revtex4-2}
\usepackage{graphicx}
\usepackage{amsmath}
\usepackage{amssymb}
\usepackage{amsfonts}
\usepackage{hyperref}
\usepackage{url}
\usepackage{color}
\usepackage{bm}

\newcommand{\be}{\begin{equation}}
\newcommand{\ee}{\end{equation}}
\newcommand{\ba}{\begin{eqnarray}}
\newcommand{\ea}{\end{eqnarray}}
\newcommand{\nn}{\nonumber}

\renewcommand{\[}{\begin{equation}}
\renewcommand{\]}{\end{equation}}

\def\lcdm{$\Lambda$CDM }

\usepackage{array}
\makeatother

\begin{document}

\preprint{IFT-UAM/CSIC-21-19 }

\title{Novel null tests for the spatial curvature and homogeneity of the Universe and their machine learning reconstructions}

\author{Rub\'{e}n Arjona}
\email{ruben.arjona@uam.es}

\author{Savvas Nesseris}
\email{savvas.nesseris@csic.es}

\affiliation{Instituto de F\'isica Te\'orica UAM-CSIC, Universidad Auton\'oma de Madrid,
Cantoblanco, 28049 Madrid, Spain}

\date{\today}

\begin{abstract}
A plethora of observational data obtained over the last couple of decades has allowed cosmology to enter into a precision era and has led to the foundation of the standard cosmological constant and cold dark matter paradigm, known as the $\Lambda$CDM model. Given the many possible extensions of this concordance model, we present here several novel consistency tests which could be used to probe for deviations from $\Lambda$CDM. First, we derive a joint consistency test for the spatial curvature  $\Omega_{k,0}$ and the matter density $\Omega_\textrm{m,0}$ parameters, constructed using only the Hubble rate $H(z)$, which can be determined directly from observations. Second, we present a new test of possible deviations from homogeneity using the combination of two datasets, either the baryon acoustic oscillation (BAO) and $H(z)$ data or the transversal and radial BAO data, while we also introduce two consistency tests for $\Lambda$CDM which could be reconstructed via the transversal and radial BAO data. We then reconstruct the aforementioned tests using the currently available data in a model independent manner using a particular machine learning approach, namely the Genetic Algorithms. Finally, we also report on a $\sim 4\sigma$ tension on the transition redshift as determined by the $H(z)$ and radial BAO data.
\end{abstract}

\maketitle


\section{Introduction \label{sec:introduction}}
Recent observations over the last couple of decades have led to overwhelming support in favor of the cosmological constant ($\Lambda$) and Cold Dark Matter (CDM) model, known as the \lcdm model \cite{Aghanim:2018eyx}. The latter significantly outperforms alternative models, as it has been noted via Bayesian analyses of astrophysical measurements \cite{Heavens:2017hkr}. However, even though the concordance, spatially flat, \lcdm model is widely accepted, some recent reanalyses of the Planck 2018 data leave the window open for a possible non-flat universe, as has been noted in Refs.~\cite{Aghanim:2018eyx,DiValentino:2019qzk,Handley:2019tkm}. This deviation from flatness could be due to unaccounted for systematic errors or due to a statistical fluctuation \cite{DiValentino:2020srs}.

In this regard, great efforts are made to provide accurate constraints on the spatial curvature of the Universe, as measured by the parameter  $\Omega_{k,0}$, since any statistically significant deviation from flatness would provide insights to the primordial inflation paradigm, aid to test physics of the early universe and also help pinpoint to high precision the age of the Universe. Moreover, accurately determining the spatial curvature of the Universe would also help in discriminating evolving dark energy density models with curvature from a flat \lcdm model, as in general evolving dark energy and curvature are degenerate with each other \cite{Virey:2008nu}. In this context several consistency tests and analyses have been proposed \cite{Clarkson:2007pz,Shafieloo:2009hi,Mortsell:2011yk,Sapone:2014nna,Rasanen:2014mca,LHuillier:2016mtc,Denissenya:2018zcv,Park:2017xbl,Cao:2021ldv,Khadka:2020tlm,Cao:2021ldv}. Actually, a detection of non-flatness, i.e $\Omega_{k,0} \neq 0$, would severely constrain the number of inflationary models, see e.g. Ref.~\cite{Guth:2013sya}, and future surveys, such as DESI and SKA, are targeting tighter measurements of $\Omega_{k,0}$ by breaking parameter degeneracies \cite{DiDio:2016ykq,Vardanyan:2009ft}.

Even though the aforementioned discrepancy might be due to unaccounted for systematic errors, there also exists the plausible possibility of new physics in the form of modified gravity (MG) or dark energy (DE) models. In fact, the $\Lambda$CDM scenario has some caveats as its main components, namely dark matter (DM) and dark energy (DE) have not yet been detected in the laboratory and are not well understood \cite{Bertone:2016nfn,Weinberg:1988cp,Carroll:2000fy}, hinting towards the idea that \lcdm could be an approximation to a more fundamental theory that remains currently unattainable \cite{DiValentino:2020vhf}. 

The existence of a large number of MG and DE models makes it difficult for the observations to be interpreted, because the results on the cosmological parameters, e.g. the value of the matter density of the Universe $\Omega_\textrm{m,0}$, depend on the particular model assumed. In fact, great efforts have recently been placed to provide a unified framework which encompasses some of these models like the Effective Field Theory (EFT) \cite{Gubitosi:2012hu,Hu:2013twa} or the Effective Fluid Approach (EFA) \cite{Arjona:2018jhh,Arjona:2019rfn,Arjona:2020gtm,Cardona:2020ama}.

One way to overcome the biases of choosing a theoretical defined model is to use non-parametric reconstruction methods and model-independent approaches \cite{Nesseris:2010ep}. In this context, machine learning (ML) algorithms have provided innovative solutions for extracting information in a theory agnostic manner \cite{Ntampaka:2019udw}. These tests are ideal to check for possible tensions that could arise because of unaccounted for systematics or could provide hints of new physics. Their main advantage is that any deviations at any redshift from the expected value imply the breakdown of any assumptions made \cite{Marra:2017pst}. Null tests have been used extensively for the concordance \lcdm model \cite{Sahni:2008xx,Zunckel:2008ti,Nesseris:2010ep}, interacting DE models \cite{vonMarttens:2018bvz}, the growth-rate data \cite{Nesseris:2014mfa,Marra:2017pst,Benisty:2020kdt}, the cosmic curvature \cite{Yahya:2013xma,Cai:2015pia,Benisty:2020otr,Li:2014yza} and also to probe the scale-independence of the growth of structure in the linear regime \cite{Franco:2019wbj}. 

Here we provide a new method to probe the spatial curvature and homogeneity of the Universe. First, we present a new joint consistency test for the curvature $\Omega_{k,0}$ and the matter density $\Omega_\textrm{m,0}$ parameters, constructed using only the Hubble rate $H(z)$, which is determined directly from observational data. This null test of the \lcdm model is an extension of the well-known $\mathcal{O}_m$ diagnostic \cite{Sahni:2008xx}, but with the added advantage that now we do not have to assume flatness. Second, we also present a null test that can be used to check for deviations from homogeneity through the combination of two datasets, either the Baryon Acoustic Oscillations (BAO) and $H(z)$ data or the transveral and radial BAO data. 

Furthermore, we also introduce two new consistency tests for \lcdm that could be tested through the transversal, also known as angular, and radial BAO respectively. The first one is derived following a similar approach to that of Ref.~\cite{Arjona:2019fwb}, where now we use the angular BAO scale relation $\theta(z)$ to present a new expression of \lcdm, which we will refer to as $\textrm{Om}_{\theta}(z)$. We show that this test has the advantage that it does not contain higher derivative terms, which increase the error when noisy data are used thus providing stringent constraints for the \lcdm model. Finally, we use the radial BAO data $\Delta z(z)$ to reconstruct the Hubble parameter $H(z)$ and the deceleration parameter $q(z)$ and constrain the accelerated expansion of the Universe.

In all cases the reconstructions of the cosmological data are performed using the Genetic Algorithms (GA), which is a stochastic minimization and symbolic regression algorithm. One of its main advantages is that it is a non parametric method which allows us to make the least number of assumptions concerning the underlying cosmology and thus avoid the issue of biases. 

The outline of our paper is as follows: in Sec.~\ref{sec:analysis} we introduce our theoretical framework. In Sec.~\ref{sec:curvature} we set out our spatial curvature and homogeneity test and in 
Sec.~\ref{sec:lcdm} we outline our \lcdm consistency tests. Then, in Sec.~\ref{sec:data} we describe the data used in our analysis and in Sec.~\ref{sec:GA} we discuss the Genetic Algorithms used to do the reconstructions. Later, in Sec.~\ref{sec:results} we present our results and in Sec.~\ref{sec:conclusions} we summarize our conclusions. Finally, in Appendix \ref{sec:appendix2} we present the results for the complementary joint null test for $\left(\Omega_m,\Omega_k\right)$ of Ref.~\cite{Seikel:2012cs}.

\section{Theoretical framework}\label{sec:analysis}
In this section we review the formalism used in the analysis and the consistency tests. Assuming that at scales of order $\sim$100Mpc the Universe is homogeneous and isotropic, then it can be described by the Friedmann-Lemaitre-Robertson-Walker (FLRW) metric at the background level, which in reduced spherical polar coordinates can be written as:
\begin{equation}\label{flrw}
 d s^{2}=-d t^{2}+a(t)^{2}\left[\frac{d r^{2}}{1-k r^{2}}+r^{2} d \theta^{2}+r^{2} \sin ^{2} \theta d \phi^{2}\right],
\end{equation}
where $t$ is the cosmic time and the scale factor $a(t)$ is related to the redshift $z$ as $a=\frac{1}{1+z}$. The spatial slices can be interpreted as flat Euclidean space with $k=0$, closed hyperspherical space with $k=+1$ or open hyperbolic space with $k=-1$. The spatial curvature of the Universe can be parameterized as $\Omega_{k,0}=-\frac{c^2}{H_0}k$, thus at late times, when we can neglect radiation since $\Omega_{r,0}\sim0$, we find that 
the Friedmann equation can be written as
\ba\label{eq:Fried}
\frac{H^{2}(z)}{H_{0}^{2}}&=& \Omega_\textrm{m,0}(1+z)^{3}+\Omega_{k,0}(1+z)^{2} \nn \\
&+&\left(1-\Omega_\textrm{m,0}-\Omega_{k,0}\right) \exp \left[3 \int_{0}^{z} \frac{1+w\left(z^{\prime}\right)}{1+z^{\prime}} d z^{\prime}\right],~~~~
\ea
where $\Omega_\textrm{m,0}$ represents the matter content of the Universe, $\Omega_{k,0}$ its curvature and $w$ the DE equation of state.
Since the cosmological constant has $w=-1$, then then  Eq.~(\ref{eq:Fried}) gives for the \lcdm model that 
\begin{equation}\label{eq:hub}
H(z)=H_{0} \sqrt{\Omega_\textrm{m,0}(1+z)^{3}+\Omega_{k,0 }(1+z)^{2}+\Omega_{\Lambda,0}},
\end{equation}
where $\Omega_{\Lambda,0}$ is related to $\Omega_\textrm{m,0}$ and $\Omega_{k,0}$ via the consistency relation
\begin{equation}
\Omega_\textrm{m,0}+\Omega_{\Lambda,0}+\Omega_{k,0}=1.
\end{equation}

The comoving distance at some redshift $z$ can be written as \cite{Weinberg:2008zzc} 
\be 
r(z)=\frac{c}{H_0}\frac1{\sqrt{-\Omega_{k,0}}}\sin \left(\sqrt{-\Omega_{k,0}}\int_0^z\frac{c}{H(z')/H_0}dz'\right),\label{eq:rz}
\ee 
while the luminosity and angular diameter distances are related via
\ba
d_L(z)&=&(1+z)\;r(z),\\
d_A(z)&=& (1+z)^{-1} r(z). \label{eq:dAeq}
\ea 
The deceleration parameter $q(z)$ is defined as
\ba\label{eq:qz}
q(z)&=&-\frac{\ddot{a}a }{\dot{a}^2}\nn \\
&=&-1+(1+z)\frac{d \ln (H/H_0)}{dz},
\ea
and assuming Eq.~(\ref{eq:hub}), at the present time ($z=0$) it can be expressed as
\ba\label{eq:q0}
q_0&\equiv& q(z=0) \nn \\
&=&\frac{1}{2}\left(-2+2\Omega_k+3\Omega_m\right).
\ea
Finally, assuming Eq.~(\ref{eq:hub}), the transition redshift $z_t$ can be defined as the redshift at which the deceleration parameter changes sign, i.e. $q(z_t)=0$. This implies that
\ba\label{eq:zt}
z_{t} &=&\left(\frac{2 \Omega_{\Lambda,0}}{\Omega_\textrm{m,0}}\right)^{1 / 3}-1 \nn\\
&=&\left(\frac{2\left(1-\Omega_\textrm{m,0}-\Omega_{k,0 }\right)}{\Omega_\textrm{m,0}}\right)^{1 / 3}-1,
\ea
which is a prediction of the \lcdm model.

\section{The null tests}\label{sec:curvature}

\subsection{Test 1: Deviations from flatness}
Defining $x=1+z$, from Eq.~(\ref{eq:hub}) we can write the matter density parameter $\Omega_\textrm{m,0}$ in terms of the Hubble function and the curvature $\Omega_{k,0}$ as
\be \label{eq:om}
\Omega_\textrm{m,0}=\frac{h^2(x)-1+\Omega_{k,0}(1-x^2)}{x^3-1},
\ee 
which reminds us of the $\mathcal{O}_m$ diagnostic of Ref.~\cite{Sahni:2008xx} when $\Omega_{k,0}\rightarrow 0$ and where we have defined $h(x)=H(x)/H_0$. The problem in this case is that now the curvature parameter, which cannot be measured in a model independent fashion, enters in the right hand side of Eq.~\eqref{eq:om}. To avoid this problem, we can use the deceleration parameter evaluated at $z=0$ given by Eq.~(\ref{eq:q0}), as it can indeed be determined independently from the data, see for example Ref.~\cite{Arjona:2019fwb}. Thus, using Eqs.~(\ref{eq:hub}) and (\ref{eq:q0}) we can simultaneously solve the algebraic system of equations for $\Omega_{k,0}$ and $\Omega_\textrm{m,0}$ to find expressions that depend on only measurements of the Hubble rate $H(z)$. Doing so we find
\ba\label{eq:om_ok1} 
\Omega_\textrm{m,0}&=&\frac{2\left(-1+h^2(z)-(1+q_0)z(2+z)\right)}{z^2\left(3+2z\right)}, \\
\Omega_{k,0}&=& \frac{3-3h^2(z)+2(1+q_0)z\left(3+z(3+z)\right)}{z^2\left(3+2z\right)},\label{eq:om_ok2}
\ea 
where $h(z)=H(z)/H_0$.

As can be seen, the joint test of Eqs.~\eqref{eq:om_ok1}-\eqref{eq:om_ok2}, is an extension of the $\mathcal{O}_m$ diagnostic of Ref.~\cite{Sahni:2008xx} as  it allows us to distinguish evolving dark energy (DE) models from the
cosmological constant, without having to assume any value for the curvature parameter. Our expressions presented here resembles that of Ref.~\cite{Seikel:2012cs}, but in our case we do not explicitly have derivatives of the Hubble rate $H(z)$, albeit only a derivative evaluated at a single point is implicitly contained in the deceleration parameter $q_0$. As we will see in later sections, this difference allows our approach to have much smaller error bars in the reconstruction compared to that of Ref.~\cite{Seikel:2012cs}.

\subsection{Test 2: Deviations from homogeneity}
Here we expand on tests of homogeneity as proposed in Ref.~\cite{Maartens:2011yx}. Homogeneity implies a  consistency relation that holds in FLRW between the angular diameter and comoving distances, given by $d_A(z)$ and $r(z)$ respectively, described by Eq.~\eqref{eq:dAeq}. Any violation of Eq.~\eqref{eq:dAeq} implies we live in a non-FLRW Universe, however, one would still expect variations on the order of $\sim10^{-5}$ due to perturbations from  large-scale structure.

One way we can test this assumption is by reconstructing separately the angular diameter distance using the BAO data and the comoving distance from the $H(z)$ data. To do so, we make use of the comoving observed BAO angle, which is given by
\begin{equation}\label{eq:thetaBAO}
    \theta_\textrm{BAO}=\frac{r_d}{(1+z)d_A(z)},
\end{equation}
and the same for the $H(z)$ data
\begin{equation}
    \theta_\textrm{H(z)}=\frac{r_d}{r(z)},
\end{equation}
where in both cases $r_d$ is the comoving sound horizon at the drag epoch $r_d \equiv r_{\rm s}(z_{\rm d})$, given by 
\be
r_{\rm s}(z_{\rm d})=\int_{z_{\rm d}}^\infty \frac{c_{\rm s}(z)}{H(z)} \,\text{d}z\, ,
\label{eq:sound-horizon-drag}
\ee
with $z_{\rm d}$ the redshift at the drag epoch, see Eq.(4) of Ref.~\cite{Eisenstein:1997ik}, while $c_{\rm s}(z)$ is the sound speed given by
\be
c_{\rm s}=\frac{c}{\sqrt{3(1+R)}},
\ee 
where $R=\frac{3 \rho_{b}}{4 \rho_{\gamma}}=\frac{3 \Omega_\mathrm{b,0}}{4 \Omega_\mathrm{\gamma,0}} a$.

Then, we can create the following expressions that can be used to search for deviations from homogeneity using BAO and $H(z)$ data: 
\ba
\zeta&=&1-\frac{\theta_\textrm{H(z)}}{\theta_\textrm{BAO}}\nn \\
&=&1-\frac{(1+z)d_A(z)}{r(z)},\label{eq:test2}
\ea
which should be zero at all $z$ for any FLRW model. 

In this case we can use BAO measurements and $H(z)$ data to directly reconstruct the angular diameter distance $d_A(z)$ and the comoving distance $r(z)$ respectively. One issue with this though is that the $H(z)$ cannot constrain the curvature parameter directly, as we reconstruct the data agnostically with the GA, thus for this test we will assume flatness, i.e. $\Omega_{k,0}=0$ in order to calculate the comoving distance $r(z)$ from the $H(z)$ data. Furthermore, had we used any data that depend on the conservation of the number of photons to measure the luminosity distance $d_L(z)$, as is for example the case for the type Ia supernovae, then the test of Eq.~\eqref{eq:test2} would in fact be a test of the cosmic distance duality (Etherington) relation $d_L(z)=(1+z)^2 d_A(z)$ instead.

We can also express Eq.~\eqref{eq:test2} using alternatively the radial $\Delta z$ and angular $\theta(z)$ BAO  data, as $d_A(z)$ and $\theta(z)$ are related via Eq.~(\ref{eq:thetaBAO}), while the radial BAO $\Delta z$ and $H(z)$ are related via the following relation
\begin{equation}\label{eq:radial}
    \Delta z=\frac{r_d \cdot H(z)}{c},
\end{equation}
where $c$ is the speed of light. Then, by using Eqs.~(\ref{eq:thetaBAO}) and (\ref{eq:radial}) we have also the following expression 
\ba\label{eq:dev_flat_2}
    \zeta=1-\frac{\theta_\textrm{H}}{\theta_\textrm{BAO}}=1-\left(\theta_\textrm{BAO}(z)\int^{z}_{0}\frac{1}{\Delta z(z')}dz'\right)^{-1},
\ea
which should be zero at all redshifts in the \lcdm model. This test has the added advantage that the radial and angular BAO are direct observables  and in fact, the sensitivity of the angular BAO scale is complementary to that of
the radial BAO \cite{Sanchez:2012eh}. 

\section{Complementary null tests\label{sec:lcdm}}

\subsection{Test 1: The angular BAO}
As mentioned before, the angular BAO can be expressed as 
\begin{equation}\label{eq:omtheta}
\theta(z,\Omega_\textrm{m,0})=\frac{r_d}{(1+z)d_A(z,\Omega_\textrm{m,0})},
\end{equation}
thus, defining the following quantity
\begin{equation}\label{eq:omtheta2}
\tilde{\theta}(z,\Omega_\textrm{m,0})=\frac{\theta}{r_d}=\frac{1}{(1+z)d_A(z,\Omega_\textrm{m,0})},
\end{equation}
we can now apply the Lagrange inversion theorem to $\tilde{\theta}(z,\Omega_\textrm{m,0})$ and write $\Omega_\textrm{m,0}$ as a function of $\tilde{\theta}(z)$, i.e $\Omega_\textrm{m,0}(z,\tilde{\theta}(z))$ via the following steps. First, in the flat \lcdm model and neglecting radiation, the angular diameter distance $d_A(z,\Omega_\textrm{m,0})$ is given by
{\small
\ba
d_A(z,\Omega_\textrm{m,0})&=&\frac{c}{H_0(1+z)} \int_0^z\frac{1}{H(x)} dx \nn\\
&=&\frac{c}{H_0}\frac{2(1+z)}{\sqrt{\Omega_\textrm{m,0}}}\left(_2F_1\left(\frac{1}{6},\frac{1}{2},\frac{7}{6},\frac{\Omega_\textrm{m,0}-1}{\Omega_\textrm{m,0}}\right)-\right.\nn\\
& &  \left. \frac{_2F_1\left(\frac{1}{6},\frac{1}{2},\frac{7}{6},\frac{\Omega_\textrm{m,0}-1}{\Omega_\textrm{m,0}(1+z)^{3}}\right)}{\sqrt{1+z}}\right).\label{eq:dl}
\ea}
Then, to derive the angular BAO test we do a series expansion on Eq.~(\ref{eq:omtheta2}) around $\Omega_\textrm{m,0}=1$ and keep the first $10$ terms in order to obtain a reliable unbiased estimation, so as to avoid theoretical systematic errors. We have chosen to keep the first $10$ terms so that at high redshifts, in particular at $z\sim2.3$ where the last of the data points are, the theoretical systematic errors are well below $\sim 1\%$.

Then, we apply the Lagrange inversion theorem to invert the series and to write the matter density $\Omega_\textrm{m,0}$ as a function of the angular BAO $\tilde{\theta}$. Then, the first two terms of the test are
\begin{equation}
\mathrm{Om}_{\theta}=1+\frac{28\left(-\frac{1}{2-2\sqrt{a}}\tilde{\theta}\right)}{\left(6+5\sqrt{a}+4a+3a^{3/2}+2a^2+a^{5/2}\right)}+O(\tilde{\theta}^2),
\end{equation}
where the scale factor $a$ is related to the redshift $z$ as $a=\frac{1}{1+z}$ and when $\tilde{\theta}$ corresponds to the \lcdm model, this should reduce to $\Omega_\textrm{m,0}$. This expression has the main advantage that it does not require taking derivatives of the data as we use the angular BAO directly and the parameter $r_d$ can also be directly obtained from the data, see Sec.\ref{sec:data} for more details.

\subsection{Test 2: The radial BAO}
In a flat \lcdm universe, the $\textrm{Om}_{H}(z)$ quantity is constant and equal to the matter energy density \cite{Sahni:2008xx}
\begin{align}
   \textrm{Om}_{H}(z) = \frac{h^2(z)-1}{(1+z)^3-1} \equiv \Omega_\textrm{m,0},
\end{align}
where $h(z)\equiv H(z)/H_0$. From Eq.~\eqref{eq:radial} we have that 
\be 
h(z)=\frac{c}{100 r_{sh}}\Delta z,
\ee
where the combination $r_{sh}=r_d \cdot h$ can be easily determined in a model independent fashion by fitting the radial BAO with the GA, see Sec.~\ref{sec:data} for more details.
Then given the relation between the Hubble parameter and the radial BAO, we can also rewrite the aforementioned expression as
\ba
\textrm{Om}_{\Delta z}(z) &=& \frac{h(z)^2-1}{(1+z)^3-1} \nn \\ 
&=& \frac{\left(\frac{c}{100r_{sh}}\right)^2\Delta z^2(z)-1}{(1+z)^3-1},
\ea

\subsection{Test 3: The deceleration parameter}
The deceleration parameter specified by Eq.~(\ref{eq:qz}) can also be estimated by using the radial BAO, see Eq.~(\ref{eq:radial}). Hence, we can measure the rate of accelerated expansion of the Universe in a model independent fashion with a different dataset other than $H(z)$ and also constrain the transition redshift $z_t$ of the acceleration phase. In this case, we can write the deceleration parameter as 
\ba\label{eq:qz2}
q(z)&=&-1+(1+z)\frac{d \ln (H/H_0)}{dz},\nn\\
&=&-1+(1+z)\frac{\Delta z'(z)}{\Delta z(z)},
\ea
and the transition redshift $z_t$ is the value at which $q(z_t)=0$. The main advantage in this case is that the radial BAO data have a much smaller error with respect to the $H(z)$ data, hence can provide stringent constraints on the deceleration parameter $q(z)$.

\section{Data}\label{sec:data}
Here we present the data we have used for our reconstructions in our analysis.

\subsection{Hubble rate data}
The Hubble rate data $H(z)$ used in our analysis is obtained by two interrelated methods. The first data set comes from the clustering of galaxies or quasars, being a direct probe of the Hubble expansion by determining the BAO peak in the radial direction \cite{Gaztanaga:2008xz}. The second compilation is obtained  by the differential age method, which is connected to the redshift drift of distant objects over long periods of time. Recall that in General Relativity (GR) the Hubble parameter can also be written in terms of the time derivative of the redshift as $H(z)=-\frac{1}{1+z}\frac{dz}{dt}$. We should note that with this last approach there are assumptions
on galaxy evolution characteristics \cite{Jimenez:2001gg}.

The $H(z)$ data used in our analysis (in units of $\textrm{km}~\textrm{s}^{-1} \textrm{Mpc}^{-1}$) comes from a compilation of $36$ points which spans a redshift range of $0.07\le z \le 2.34$ and is based on those of Refs.~\cite{Moresco:2016mzx,Zhang:2012mp,STERN:2009EP,MORESCO:2012JH,Chuang:2012qt,Blake:2012pj,Anderson:2013zyy,Delubac:2014aqe} and Ref.~\cite{Guo:2015gpa}. The compilation, which can be found at Table I of Ref.~\cite{Arjona:2018jhh}, comes in the form $(z_i,H_i,\sigma_{H_i})$. We have minimized the $\chi^2$ analytically over $H_0$ finding
\ba
\chi^2_{H}&=&A-\frac{B^2}{\Gamma},\label{eq:chi2H}\\
H_0&=&\frac{B}{\Gamma},\label{eq:H0bf}
\ea
where the parameters $A$, $B$ and $\Gamma$ are defined as
\ba
A&=&\sum_i^{N_H}\left(\frac{H_i}{\sigma_{H_i}}\right)^2, \\
B&=&\sum_i^{N_H}\frac{H_i~E^{th}(z_i)}{\sigma_{H_i}^2}, \\ \Gamma&=&\sum_i^{N_H}\left(\frac{E^{th}(z_i)}{\sigma_{H_i}}\right)^2,
\ea
and we designate the theoretical value of the Hubble parameter as $E^{th}(z)=H^{th}(z)/H_0$, while the number of points is $N_H=36$.

This data set has been used to extract different cosmological information such as measuring the Hubble constant $H_0$, determine the deceleration transition redshift, constrain the spatial curvature of the Universe along with distance redshift data and also the non-relativistic matter and DE parameters as can be seen in \cite{Yu:2017iju}.

\subsection{BAO data}
The different BAO data used in our analysis comes from 6dFGS \cite{Beutler:2011hx}, SDDS \cite{Anderson:2013zyy}, BOSS CMASS \cite{Xu:2012hg}, WiggleZ \cite{Blake:2012pj}, MGS \cite{Ross:2014qpa} and BOSS DR12 \cite{Gil-Marin:2015nqa}, DES \cite{Abbott:2017wcz}, Lya \cite{Blomqvist:2019rah}, DR - 14 LRG \cite{Bautista:2017wwp} and quasars \cite{Ata:2017dya}. The following functions that we will present now are used to describe the data. First, we define the ratio of the sound horizon at the drag redshift to the so called dilation scale:  
\be
d_z\equiv \frac{r_s(z_d)}{D_V(z)},\label{eq:dz}
\ee
where the sound horizon is given by Eq.~\eqref{eq:sound-horizon-drag} and $z_d$ is the redshift at the dragging epoch, see Eq.~(4) of \cite{Eisenstein:1997ik}. In the \lcdm model the sound horizon can be approximated as
\be
r_s(z_d)\simeq \frac{44.5\log\left(\frac {9.83} {\Omega_\textrm{m,0}h^2}\right)}{\sqrt {1+10(\Omega_\textrm{b,0}h^2)^{3/4}}} \textrm{Mpc},\label{eq:rd}
\ee
while the dilation scale is given by
\be
D_V(z)=\left[(1+z)^2 d_A(z)^2 \frac{c z}{H(z)}\right]^{1/3},
\ee
where we have defined the Hubble distance
\be
D_H(z)=c/H(z).
\ee
Then, the 6dFGs and WiggleZ BAO data are specified as
\be
\begin{array}{ccc}
 z  & d_z & \sigma_{d_z } \\
 \hline
 0.106 & 0.336 & 0.015 \\
 0.44 & 0.073 & 0.031 \\
 0.6 & 0.0726 & 0.0164 \\
 0.73 & 0.0592 & 0.0185 \\
\end{array}
\ee
where their inverse covariance matrix is
\be C_{ij}^{-1}=\left(
\begin{array}{cccc}
 \frac{1}{0.015^2} & 0 & 0 & 0 \\
 0 & 1040.3 & -807.5 & 336.8 \\
 0 & -807.5 & 3720.3 & -1551.9 \\
 0 & 336.8 & -1551.9 & 2914.9 \\
\end{array}
\right)\ee
with the $\chi^2$ given by
\be
\chi^2_\textrm{6dFS,Wig}=V^i C_{ij}^{-1} V^j,\label{eq:chi26df}
\ee
and $V^i=d_{z,i}-d_z(z_i,\Omega_\textrm{m,0})$.

The BAO measurements for MGS and SDSS (LowZ and CMASS) are given by $D_V/r_s = 1/d_z$ via
\be\begin{array}{ccc}
 z  & 1/d_z & \sigma_{1/d_z } \\
 \hline
 0.15 & 4.46567 & 0.168135 \\
 0.32 & 8.62 & 0.15 \\
 0.57 & 13.7 & 0.12 \\
\end{array}\ee
and the
\be
\chi^2_\textrm{MGS,SDSS}=\sum \left(\frac{1/d_{z,i}-1/d_z(z_i,\Omega_\textrm{m,0})}{\sigma_{1/d_{z,i}}}\right)^2.\label{eq:chi2SDSS}
\ee
At this point we should stress that these aforementioned data points were provided by their respective collaborations, 6dFGs and WiggleZ for the ones in Eq.~\eqref{eq:chi26df} and MGS and SDSS for Eq.~\eqref{eq:chi2SDSS} in that exact form as this is how they are extracted from the raw data. Hence, we have not made any assumptions from our part at this stage. 

The BAO data from DES are of the form $d_A(z)/r_s$ with $(z,d_A(z)/r_s,\sigma)=(0.81, 10.75, 0.43)$ and the $\chi^2$ given by \be
\chi^2_\textrm{DES}=\sum \left(\frac{d_A({z,i})/r_s-d_A(z_i,\Omega_\textrm{m,0})/r_s}{\sigma_{d_A(z,i)/r_s}}\right)^2.
\ee
We also include the BAO data from Lya, which are of the form $f_\textrm{BAO}=((1+z)d_A/r_s, D_H/r_s)$ and are given by
\be
\begin{array}{ccc}
 z & f_\textrm{BAO}  &  \sigma_{f_\textrm{BAO}} \\
 \hline
 2.35 & 36.3 & 1.8 \\
 2.35 & 9.2 & 0.36 \\
\end{array}
\ee
with the $\chi^2$ given by
\be
\chi^2_\textrm{Lya}=\sum \left(\frac{f_\textrm{BAO,i}-f_\textrm{BAO}(z_i,\Omega_\textrm{m,0})}{\sigma_{f_\textrm{BAO}}}\right)^2.
\ee

Finally, the DR-14 LRG and quasars BAO data make the assumption of $r_{s,fid} = 147.78~\textrm{Mpc/h}$ and are given by $D_V/r_s = 1/d_z$
\be
\begin{array}{ccc}
 z  & 1/d_z & \sigma_{1/d_z } \\
 \hline
 0.72 & \frac{2353}{r_{s,fid}} & \frac{62}{r_{s,fid}} \\
 1.52 & \frac{3843}{r_{s,fid}} & \frac{147}{r_{s,fid}} \\
\end{array}
\ee
and the $\chi^2$ given by
\be
\chi^2_\textrm{LRG,Q}=\sum \left(\frac{1/d_{z,i}-1/d_z(z_i,\Omega_\textrm{m,0})}{\sigma_{1/d_{z,i}}}\right)^2.
\ee
The total $\chi^2$ is then given by
\be
\chi^2_\textrm{tot}=\chi^2_\textrm{6dFS,Wig}+ \chi^2_\textrm{MGS,SDSS} + \chi^2_\textrm{DES} +\chi^2_\textrm{Lya}+\chi^2_\textrm{LRG,Q}.\label{eq:chi2tot}
\ee
Note that the previous equation carries the assumption that the data are independent, hence we can just add the $\chi^2$ together. As some of the data points are from the same survey, there must be galaxies in common between the overlapping datasets, and therefore some potentially strong covariances, something which poses an important limitation of our analysis. 

In the particular cases, e.g. the WiggleZ data, where the correlations between the points, quantified in terms of a covariance matrix $C_{ij}$, are known, we have then included the $C_{ij}$ in our analysis. However, in most cases the full correlations are in practice not publicly available or it is impossible to correctly estimate a covariance matrix, even if a few attempts have been made in the literature, e.g. for a similar discussion for the $f\sigma8$ data see Ref.~\cite{Alam:2015rsa}. 

One way to resolve this important issue was proposed in Ref.~\cite{Alam:2015rsa}, where the authors approximated the overall covariance matrix of the $f\sigma8$ measurements as the percent fraction of overlapping volume between the surveys to the total volume of the two surveys combined. However, clearly this approach cannot take into account any negative correlations between the data as in general, the effect of the correlations can also be due to instrument systematics etc. Thus, approximating the covariance matrix with the percent overlap can potentially lead to a strongly biased covariance matrix. 

On a related note, another limitation of this compilation of the BAO data is that some of the data points, e.g. those coming from 6dFGs and  WiggleZ, are given in terms of the variable $d_z(z)$, while some other points, e.g. from MGS and SDSS, are given in terms of $1/d_z(z)$. This poses a problem as we have to make an implicit assumption about how the errors in those data points are distributed. For example, we always assume symmetric error bars, but if the errors are symmetric in $1/d_z(z)$, they will not be symmetric in $d_z(z)$ and vice versa. This further complicates the analysis as $d_z(z)$ and $1/d_z(z)$ are not raw data that can be reexpressed in a more consistent form, but they are derived data products which makes it impossible to rewrite them in the same form, without making further assumptions.

Finally, the BAO $\chi^2$ terms in Eq.~\eqref{eq:chi2tot} depend on the sound speed at the drag redshift $r_d=r_s(z_d)$ through Eq.~\eqref{eq:dz}, which is complicated to estimate model independently. In order to not assume a value for $H_0$ in our reconstruction when we fit the BAO data we have minimized the $\chi^2$ over the quantity $r_{sh}=r_s\cdot h$, where $r_s$ is the sound horizon at the drag redshift and $h$ is the Hubble parameter. Hence, we avoid any bias of the results due to specifying a value of $H_0$.

\subsection{Radial BAO data}
The 6 data points for the radial BAO $\Delta z$ are taken from Table III of the SDSS-IV spectroscopic survey \cite{Alam:2020sor} coming from SDSS, SDSS-II, BOSS and eBOSS. As we did with the BAO data, to do not assume a value for $H_0$ in our reconstruction when we fit the radial BAO data we have minimized the $\chi^2$ over the quantity $r_{sh}=r_s\cdot h$.

\subsection{Angular BAO data}
The angular BAO, also known as the transversal BAO scale data have been taken from Table I of Ref.~\cite{Nunes:2020hzy} where 15 measurements of $\theta(z)$ are given and where the data have been derived without assuming a fiducial cosmology, following the approach of Ref.~\cite{Sanchez:2010zg}. In particular, the compilation of the angular BAO data comes from luminous red galaxies, blue galaxies, quasar catalogs and from diverse releases of the Sloan Digital Sky Survey (SDSS), see Refs.~\cite{deCarvalho:2017xye, Carvalho:2017tuu, Carvalho:2015ica, Alcaniz:2016ryy}. For the fit of the transverse BAO we have also minimized the $\chi^2$ over the quantity $r_{sh}=r_s\cdot h$. 
\section{Genetic Algorithms\label{sec:GA}}
The GA are a stochastic optimization machine learning approach that can be used for non-parametric reconstruction of a given data set. The fundamental principle of the GA is very loosely related to natural selection, where the species evolve over the aeons due to evolutionary pressure of this natural process. In our case our population under study would represent a set of test functions that are going to evolve and change over time through the stochastic operations of mutation and crossover. The former refers to the combination of different individuals to produce offspring and the latter to a random swap in the chromosomes of an individual. Then the strategy of the GA is to find an analytical function that represent the data employing on or more variables.

In our analysis we use the definition of the $\chi^2$ statistic to specify how well each individual agrees with the data set. Then the probability that a population of functions will produce offspring will be proportional to its fitness defined by this $\chi^2$. The data set used in the analysis is described in Sec.~\ref{sec:data}, where the $H(z)$ compilation, the BAO data and the angular and radial BAO data are used to reconstruct the Hubble rate $H(z)$, the angular diameter distance $d_A(z)$, the angular BAO $\theta(z)$ and the radial BAO $\Delta z(z)$ respectively.
 
To perform the reconstructions in our analysis we implemented the following approach. First, our grammar included the following orthogonal basis of functions: exp, log and polynomials and a set of operations $+,-,\times,\div, \wedge$. The choice of the grammar and the size of the population has been extensively tested in Ref.~\cite{Bogdanos:2009ib} finding that it does influence the convergence rate of the GA. We also specified some assumptions motivated by physical reasons. For instance at the present day $z=0$ we have that $H(z=0)=H_0$, $d_A(z=0)=0$ and similarly $\theta(z=0)\sim \frac{rsh}{z}$ and $\Delta z(z=0)\sim \frac{100rsh}{c}$, but we make no assumptions on the curvature of the Universe or any MG or DE model. We also imposed that all the functions the GA reconstructs are continuous and differentiable, without any singularities in the redshift scanned by the data to avoid overfitting or fake reconstructions.

The GA will initialize with a random population, i.e a set of functions whose size is a heuristic parameter. When the starting population has been built, the fitness of each individual is evaluated by a $\chi^2$ statistic. Afterwards, choosing the candidates for crossover by considering smaller groups of randomly selected individuals and selecting the dominant member of each group, see  Ref.~\cite{Bogdanos:2009ib}, a process known as tournament selection, the best-fitting functions in every generation are selected and the operations of crossover and mutation are used. To warranty convergence, the GA code is repeated thousands of times and exploring different random seeds, in order to rightly inspect the functional space. The final output of the code, then is a set of smooth and analytic functions for $H(z)$, $d_A(z)$ and $\theta(z)$, $\Delta z(z)$ that describe the data.

\begin{figure*}[!t]
\centering
\includegraphics[width = 0.48\textwidth]{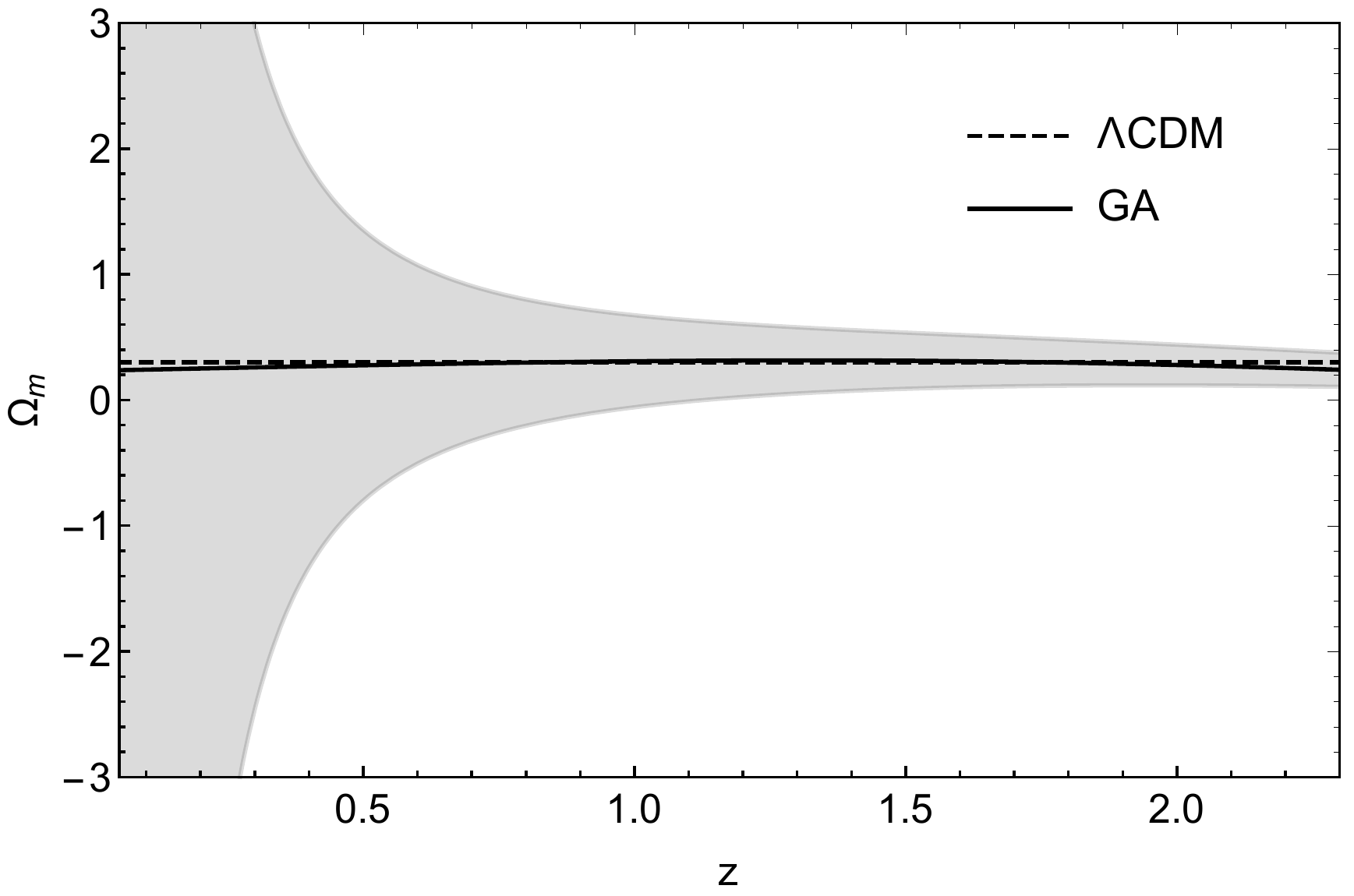}
\includegraphics[width = 0.48\textwidth]{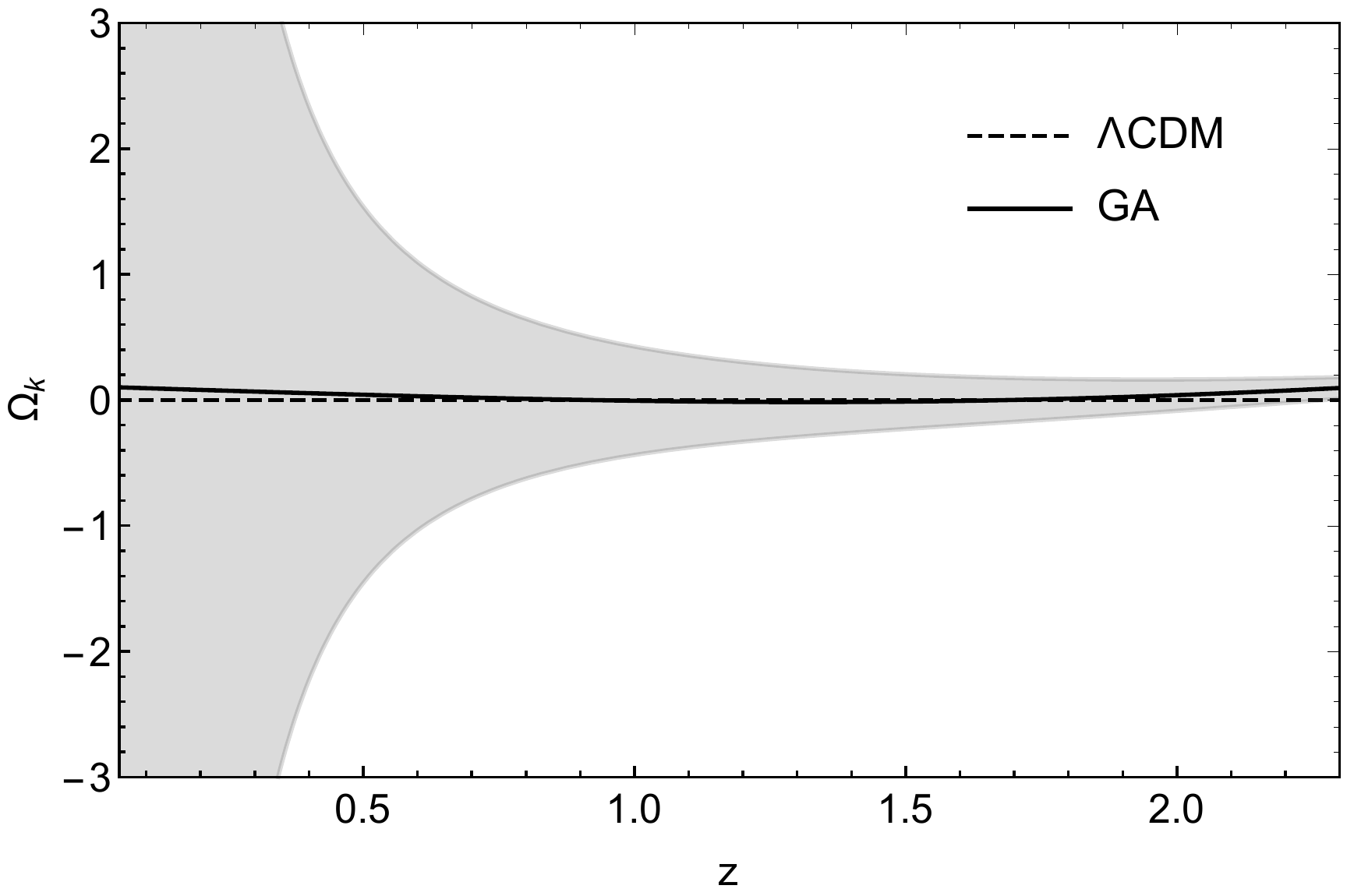}
\caption{The GA reconstruction of the  $\Omega_m(H,q_0)$ (left panel) and $\Omega_k(H,q_0)$ (right panel) expressions given by Eqs.~\eqref{eq:om_ok1} and \eqref{eq:om_ok2} respectively and obtained using the $H(z)$ data. In both cases the black solid line and the grey region corresponds to the GA best-fit and the $1\sigma$ error respectively. As can be seen, both reconstructions are consistent with the flat \lcdm model represented by the black dashed-line at the $1\sigma$ level. \label{fig:ok_H}}
\end{figure*}

\begin{figure*}[!t]
\centering
\includegraphics[width = 0.48\textwidth]{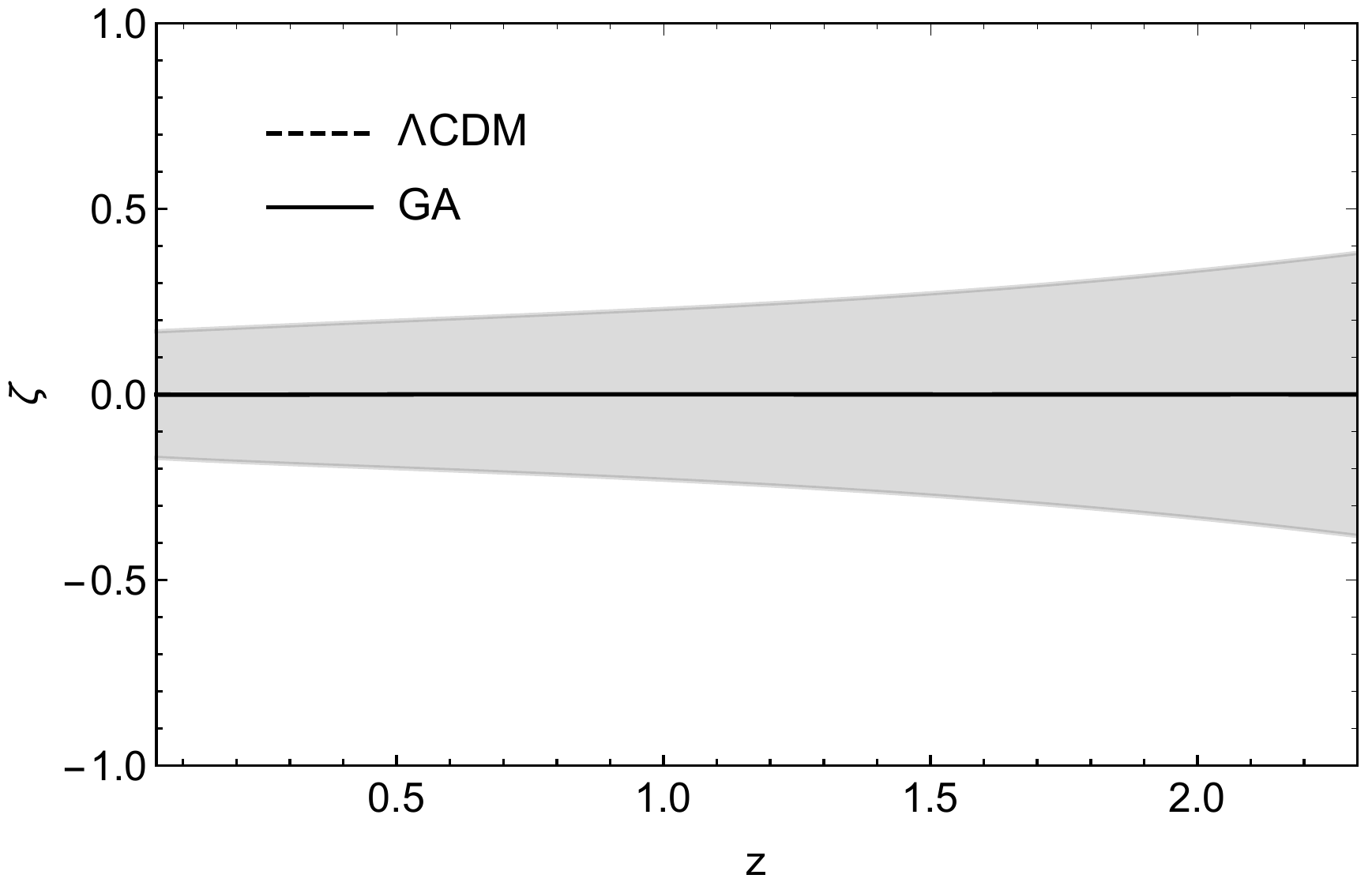}
\includegraphics[width = 0.48\textwidth]{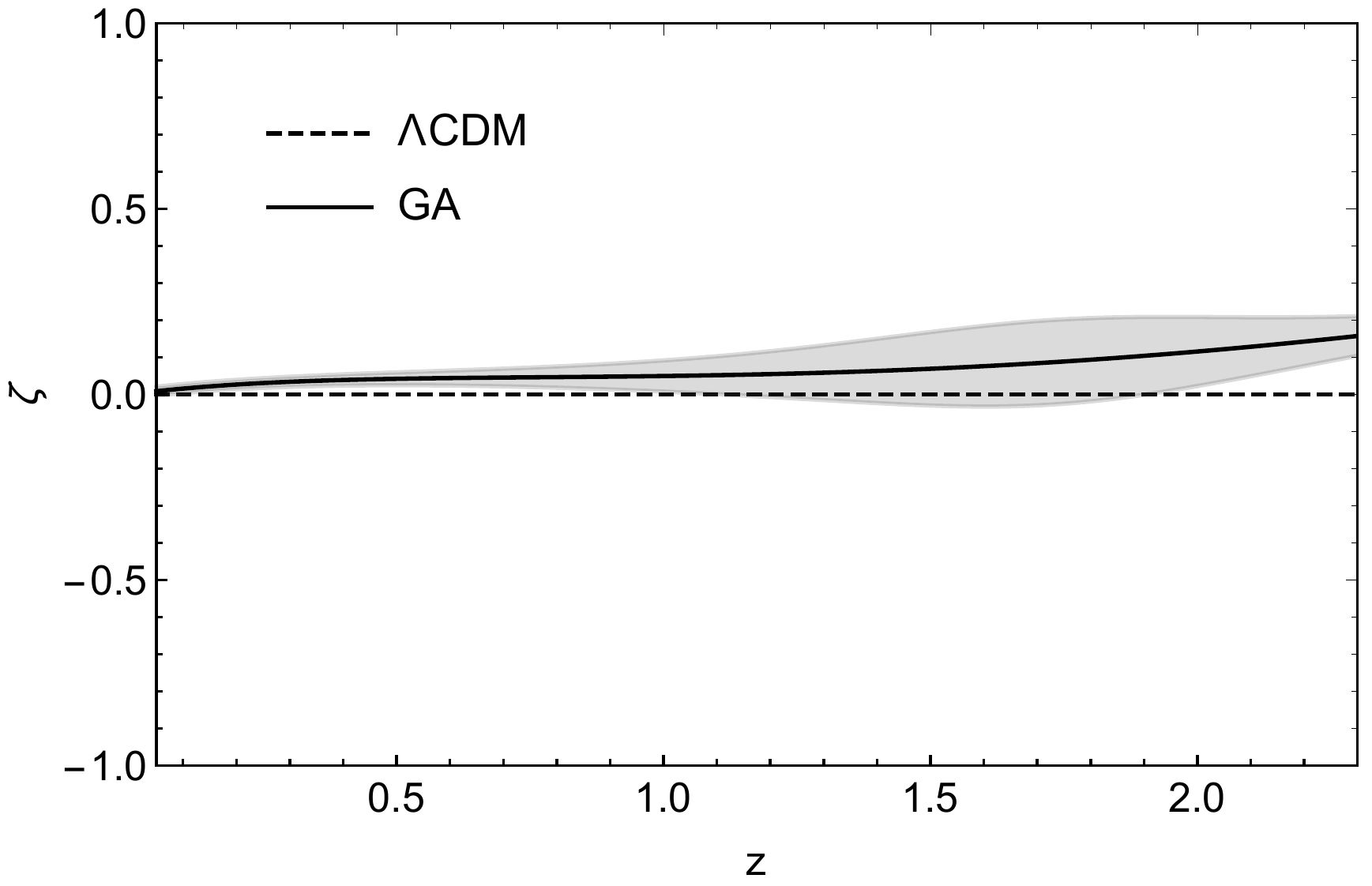}
\caption{The $\zeta=1-\frac{\theta_{H}}{\theta_\textrm{BAO}}$ test which aims to find deviations from homogeneity through our reconstructions from the $H(z)$ and BAO data (left panel) and our angular and radial BAO data (right panel). In both cases the black solid line and the grey region corresponds to the GA best-fit and the $1\sigma$ error respectively. As can be seen, both reconstructions are consistent with the expectation of no deviation, represented by the black dashed-line, at the $1\sigma$ level in the left and at the $\sim 2\sigma$ in the right. 
\label{fig:deviation}}
\end{figure*}

Concerning the errors of the reconstructed functions, they are obtained through a method originally implemented in Refs.~\cite{Nesseris:2012tt,Nesseris:2013bia} known as the path integral approach. It consists of estimating the error regions by integrating analytically the likelihood over all possible functions that might be constructed by the GA. This error reconstruction method, which results in Gaussian errors, has been intensely examined and compared against a bootstrap Monte Carlo simulation by Ref.~\cite{Nesseris:2012tt}.

A discussion on the implementation of the Genetic Algorithms (GA) can also be found on Sec.~(4) of \cite{Arjona:2020skf}, Sec.~(C.1) of \cite{Arjona:2020axn}, Appendix~(C) of \cite{Arjona:2020doi}, Sec.~(4) of \cite{Arjona:2020kco} and Sec.~(2) of \cite{Arjona:2019fwb} among others. For recent applications of the GA on Cosmology see Refs.~\cite{Bogdanos:2009ib,Nesseris:2012tt,Arjona:2020doi,Arjona:2020kco,Arjona:2019fwb,Arjona:2020skf,Arjona:2020axn,Nayak:2021lzf}. 

\begin{figure*}[!thp]
\centering
\includegraphics[width = 0.48\textwidth]{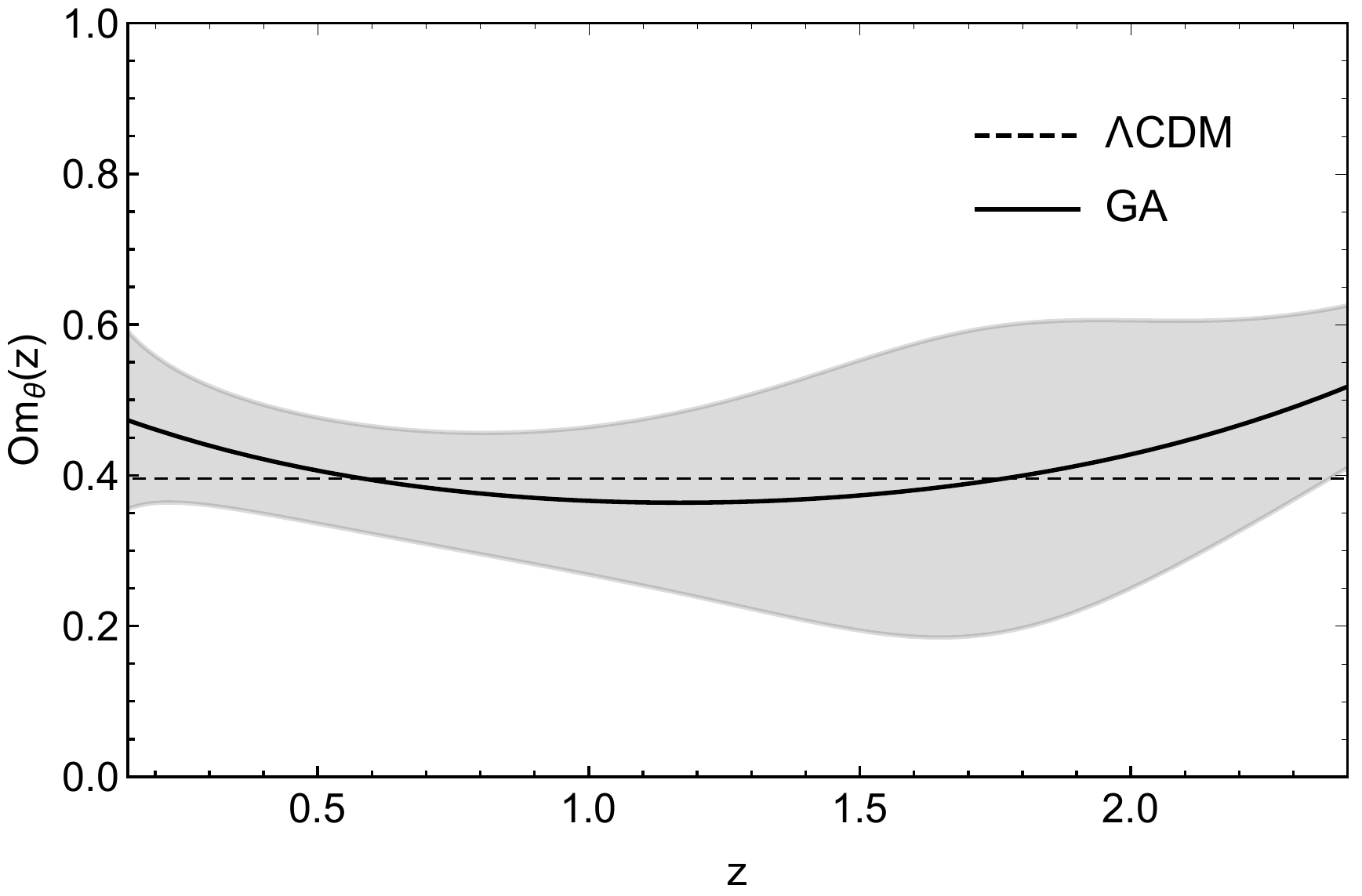}
\includegraphics[width = 0.48\textwidth]{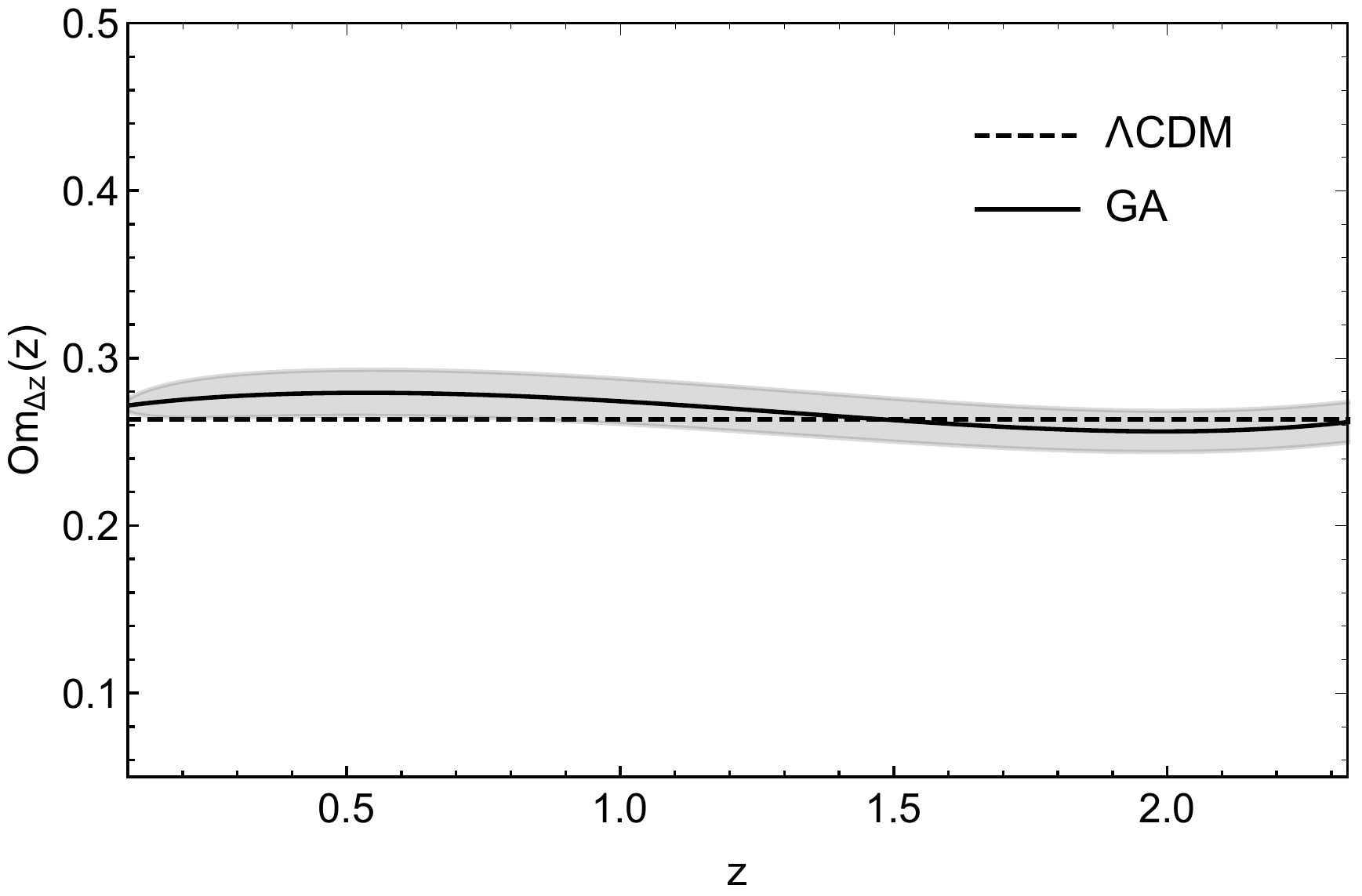}
\caption{Left panel: The reconstruction of the $\text{Om}_{\theta}(z)$ statistics obtained through our reconstruction of the angular BAO data using the GA. Right panel:  The reconstruction of the $\text{Om}_{\Delta z}(z)$ statistics derived through our GA reconstruction of the radial BAO data. In both cases the black solid line and the grey region corresponds to the GA best-fit and the $1\sigma$ error respectively. Both reconstructions are consistent with the best-fit flat \lcdm model represented by the black dashed-line at the $1\sigma$ level.\label{fig:omtest}}
\end{figure*}

\begin{figure}[!htp]
\centering
\includegraphics[width = 0.48\textwidth]{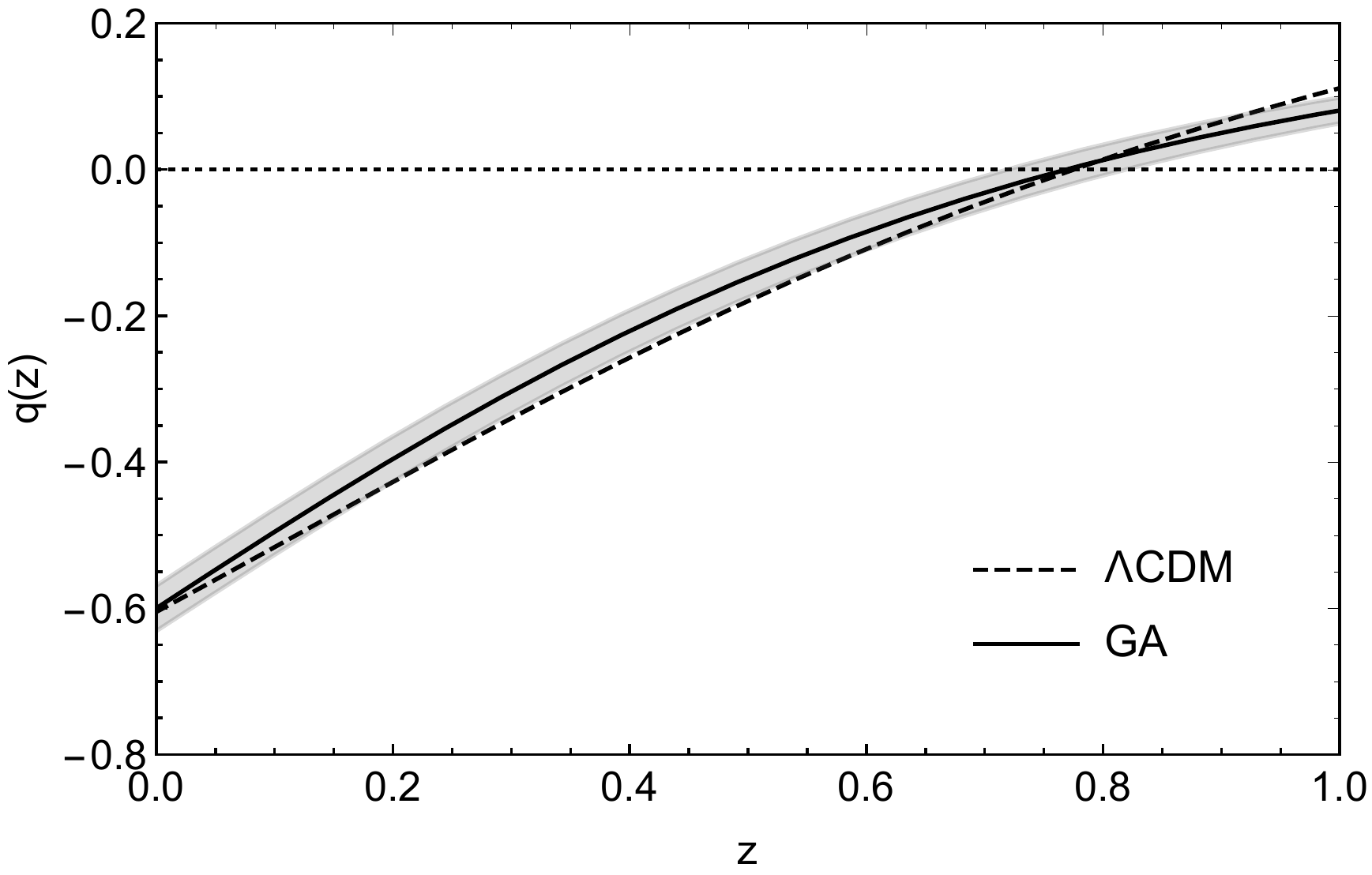}
\caption{The deceleration parameter given by Eq.~(\ref{eq:qz}) as reconstructed by using Eq.~(\ref{eq:radial}). The black solid line and the grey region corresponds to the GA best-fit and the $1\sigma$ error respectively. Our model independent detection of the accelerated expansion of the Universe is consistent with the best-fit flat \lcdm model represented by the black dashed-line at the $1\sigma$ level. The transition redshift $z_{tr}$ corresponds to the point where $q(z)$ crosses zero. \label{fig:radial_test}}
\end{figure}

\section{Results}\label{sec:results}
In this section we present our GA fits to the data and the corresponding consistency tests derived through our reconstructions. In Table~\ref{tab:GAchi2} we show the best-fit $\chi^2$ per degree of freedom (dof) or equivalently per number of points, for the GA functions and the best-fit \lcdm model. As can be seen, in all cases the GA out-performs the \lcdm model in terms of the best-fit $\chi^2/\textrm{dof}$.

Concerning our probe for the curvature, in Fig.~\ref{fig:ok_H} we present the $\Omega_m(H,q_0)$ (left panel) and $\Omega_k(H,q_0)$ (right panel) expressions given by Eqs.~\eqref{eq:om_ok1} and \eqref{eq:om_ok2} respectively, obtained through our GA reconstruction of the $H(z)$ data.\footnote{In Appendix \ref{sec:appendix2} we also present a complementary test for $\Omega_k$ containing derivatives of $H(z)$.} In both cases the black solid line and the grey region correspond to the GA best-fit and its $1\sigma$ error respectively. As can be seen, both reconstructions are consistent with the best-fit flat \lcdm model, represented by the black dashed-line, at the $1\sigma$ level.

In Fig.~\ref{fig:deviation} we show the $\zeta=1-\frac{\theta_{H}}{\theta_{BAO}}$ test which aims to find deviations from homogeneity through our reconstructions from the $H(z)$ and BAO data (left panel) and our angular and radial BAO data (right panel). In both cases the black solid line and the grey region corresponds to the GA best-fit and the $1\sigma$ error respectively. As can be seen, both reconstructions are consistent with the best-fit flat \lcdm model, represented by the black dashed-line, at the $1\sigma$ level in the left and at $\sim 2\sigma$ in the right panel.

\begin{table}[]
    \centering
     \begin{tabular}{ccccc} 
 & $H(z)$ & BAO & $\theta(z)$ & $\Delta z(z)$ \\
\hline$\chi_{\Lambda \mathrm{CDM}}^{2}/\textrm{dof}$ & 0.541 &   0.911 & 0.843 & 0.734 \\
\hline$\chi_\textrm{GA}^{2}/\textrm{dof}$ & 0.491 & 0.610 & 0.831 & 0.592 \\
\hline
\end{tabular}
 \caption{The $\chi^2/\textrm{dof}$ for \lcdm and GA using the Hubble rate $H(z)$, the BAO data and the angular $\theta(z)$ and radial $\Delta z(z)$ BAO data.}
    \label{tab:GAchi2}
\end{table}

In Fig.~\ref{fig:omtest} we show our consistency tests of the \lcdm model. In particular, in the left panel we show the reconstruction of the $\text{Om}_{\theta}(z)$ statistic, obtained through our reconstruction of the angular BAO data using the GA. On the right panel we have the reconstruction of the $\text{Om}_{\Delta z}(z)$ statistics derived through our GA reconstruction of the radial BAO data. In both cases the black solid line and the grey region corresponds to the GA best-fit and the $1\sigma$ error respectively. Both reconstructions are consistent with the best-fit flat \lcdm model, represented by the black dashed-line, at the $1\sigma$ level. It is worth noting that the best-fit value of the matter density for the flat \lcdm model is given by $\Omega_{m,0}=0.396\pm 0.154$, which is somewhat higher than the one found by other observations \cite{Aghanim:2018eyx}. A possible explanation for this higher value of the matter density could be due to the assumptions made on Section~\ref{sec:curvature}, where we are reducing the complex galaxy survey data to single values of $\theta(z)$.

Furthermore, in Fig.~\ref{fig:radial_test} we present the deceleration parameter given by Eq.~(\ref{eq:qz}) as reconstructed by using Eq.~(\ref{eq:radial}). The black solid line and the grey region corresponds to the GA best-fit and the $1\sigma$ error respectively. Our model independent detection of the accelerated expansion of the Universe is consistent with the best-fit flat \lcdm model, represented by the black dashed-line, at the $1\sigma$ level. The transition redshift $z_{tr}$ corresponds to the point where $q(z)$ crosses zero, and $q(z)$ is obtained via Eq.~(\ref{eq:qz2}).

Finally, with our GA reconstructions we find the following derived parameters
\ba
r_{s}(\text{BAO})&=&101.873\pm 2.078\hspace{2mm} \text{Mpc/h},\\
r_{s}(\text{transverse BAO})&=&103.938\pm 2.132\hspace{2mm} \text{Mpc/h},\\
r_{s}(\text{radial BAO})&=&103.477\pm 1.447\hspace{2mm} \text{Mpc/h},
\ea
while from the radial BAO we also find
\ba
q_\textrm{GA,0}&=&-0.600\pm 0.031,\\
z_\textrm{GA,tr}&=&0.769\pm 0.050,
\ea
where $z_\textrm{tr}$ is the value of the transition redshift, i.e the moment when the deceleration parameter changes sign. It should be noted that using $H(z)$ data, Ref.~\cite{Arjona:2019fwb} had reported a value for the deceleration parameter today of $q_0=-0.575\pm 0.132$ and the transition redshift $z_{tr} = 0.662 \pm 0.027$, where the latter is $\sim 4\sigma$ away from the value reported earlier, thus hinting at a possible tension between the two datasets.

In this case, our constraint of the transition redshift using the $H(z)$ data and the radial BAO data comes from the same method, the GA. Actually, the main cause of the difference is due to the larger errors of the $H(z)$ data, and the associated possible systematics in the cosmic chronometers, compared to the errors of the radial BAO. This difference causes a small difference between the best-fit value from \lcdm and the GA for the $H(z)$ data, while the radial BAO the GA and the best-fit value of \lcdm are more in agreement. Hence, this points to possible issues with the data, even if they have a small overlap with some of the points.

\begin{figure*}[!thp]
\centering
\includegraphics[width = 0.48\textwidth]{img/ok_H.pdf}
\includegraphics[width = 0.48\textwidth]{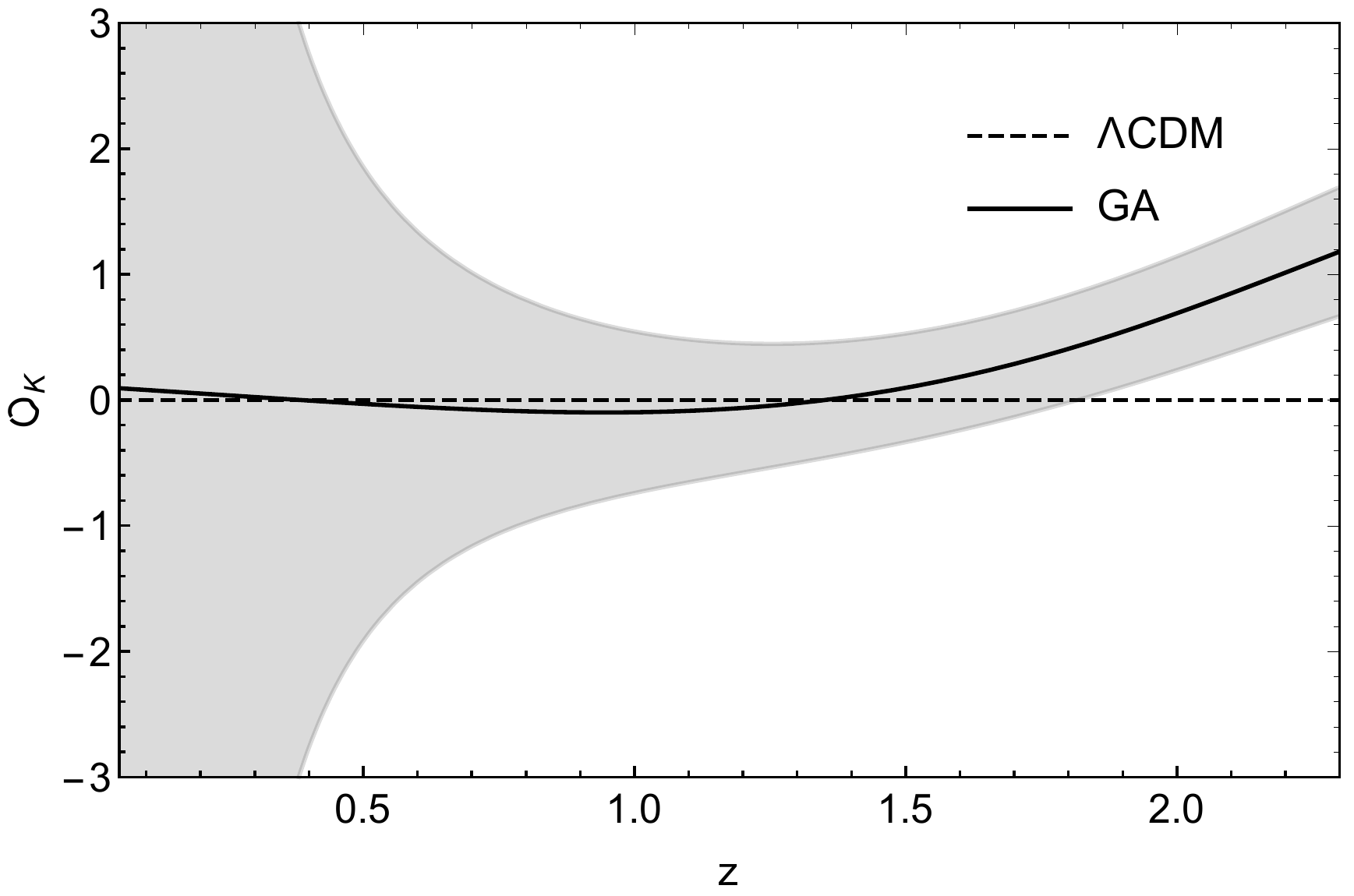}
\caption{The $\Omega_k(H,q_0)$ expression presented in this paper (left panel) compared to the $\mathcal{O}_{K}(z)$ expression from Ref.~\cite{Seikel:2012cs} (right panel). In both cases we use our GA reconstruction of the $H(z)$ data, while the black solid line and the grey shaded regions correspond to the GA best-fit and the $1\sigma$ error respectively.  \label{fig:ok_H_comp}}
\end{figure*}

\section{Conclusions}\label{sec:conclusions}
In this work we have presented a set of new consistency tests for the spatial curvature and homogeneity of the Universe and the \lcdm model, which is the target of upcoming large-scale structure surveys. These tests will provide us alternative and complementary tests of the validity of the standard flat $\Lambda$CDM paradigm. 

In our analysis we prefer to use the GA, compared to other non-parametric approaches, e.g. various kinds of orthogonal polynomials or Gaussian Processes. The reason is that the Gaussian Processes still require the choice of a mean function, arbitrarily assumed to be either some constant, e.g. zero, or the cosmological constant \lcdm model. In previous works we have shown that the GA do not suffer from this issue, hence we believe they are quite appropriate for the problem at hand.  

In our work we presented an extension of the consistency diagnostic of flat \lcdm of Ref.~\cite{Sahni:2008xx}, by now including both the curvature $\Omega_{k,0}$ and the matter density $\Omega_\textrm{m,0}$, see Eqs.~(\ref{eq:om_ok1})-(\ref{eq:om_ok2}). We show how the latter can be derived from the Hubble rate $H(z)$ and can be determined directly from observational data. We should stress that the added advantage of our new null test of the \lcdm model presented here is the fact that we do not have to assume a flat Universe.

Secondly, through the combination of the BAO and $H(z)$ data or the angular and radial BAO data respectively, we also presented a test to search for deviations from homogeneity, see Eqs.~(\ref{eq:test2}) and (\ref{eq:dev_flat_2}). Then, with the angular and radial BAO data we also introduce two new consistency tests for the \lcdm model. The first one, namely $\textrm{Om}_{\theta}(z)$, is derived following a similar approach as it was shown in Ref.~\cite{Arjona:2019fwb} where in this case we use the angular BAO scale relation $\theta(z)$ to reconstruct null tests of the \lcdm model, with the advantage that this null test does not contain higher derivative terms, which tend to increase the reconstruction errors when using noisy data.

Finally, we also used the radial BAO data $\Delta z(z)$ to obtain a model independent determination of the accelerated expansion of the Universe by reconstructing the deceleration parameter $q(z)$ and we applied our $\Delta z$ reconstruction to the $\mathcal{O}_m$ diagnostic \cite{Sahni:2008xx}. Overall we find that our results are consistent with the standard flat $\Lambda$CDM scenario, however we also noted a $\sim 4 \sigma$ tension on the determination of the transition redshift $z_t$, i.e. the redshift where the Universe transitions from decelerated to accelerated expansion, between the $H(z)$ and the radial BAO data. 

Overall, we find that these new tests of the spatial curvature and homogeneity of the Universe can be used, in a model-independent fashion, to test some of the fundamental assumptions of the standard cosmological model. However, our reconstructions are somewhat limited by the current data, albeit this should be resolved in the near future when high quality BAO data become available from the next stage surveys.\\

\section*{Acknowledgements}
The authors thank F. Javier Sánchez for useful discussions and acknowledge support from the Research Project  PGC2018-094773-B-C32 and the Centro de Excelencia Severo Ochoa Program SEV-2016-0597. S.~N. also acknowledges support from the Ram\'{o}n y Cajal program through Grant No. RYC-2014-15843.\\

Numerical Analysis Files: The Genetic Algorithm code used by the authors in the analysis of the paper can be found at  \href{https://github.com/RubenArjona}{https://github.com/RubenArjona}.\\

\begin{appendix} 

\section{Complementary null tests \label{sec:appendix2}}
Here we briefly present a complementary $\left(\Omega_{m,0},\Omega_{k,0}\right)$ joint test with a similar one from Ref.~\cite{Seikel:2012cs}, which is defined as

\ba
\mathcal{O}_{m}^{(2)}(z) & \equiv& 2 \frac{(1+z)\left(1-h^{2}\right)+z(2+z) h h^{\prime}}{z^{2}(1+z)(3+z)}, \\
\mathcal{O}_{K}(z) & \equiv & \frac{3(1+z)^{2}\left(h^{2}-1\right)-2 z\left(3+3 z+z^{2}\right) h h^{\prime}}{z^{2}(1+z)(3+z)},~~~~~~\label{eq:CKok}
\ea
where $h=H(z)/H_0$ and the prime $'$ is a derivative with respect to $z$. The previous tests in the \lcdm limit reduce to
\ba
\mathcal{O}_{m}^{(2)}(z)&=&\Omega_{m,0},  \\
 \mathcal{O}_{K}(z)&=&\Omega_{k,0}.
\ea
In Fig.~\ref{fig:ok_H_comp} we show the two reconstructions for the curvature test, on the left panel for our test given by Eq.~\eqref{eq:om_ok2} and in the right panel for that of Ref.~\cite{Seikel:2012cs} given by Eq.~\eqref{eq:CKok}. The main difference of the latter with our test comes at high redshifts $z>1$, where the estimated errors of the GA are smaller with our test and also agree more at high redshifts with the expectation from Planck of a nearly flat Universe. 
\end{appendix}

\bibliography{curvature}

\begin{thebibliography}{84}%
\makeatletter
\providecommand \@ifxundefined [1]{%
 \@ifx{#1\undefined}
}%
\providecommand \@ifnum [1]{%
 \ifnum #1\expandafter \@firstoftwo
 \else \expandafter \@secondoftwo
 \fi
}%
\providecommand \@ifx [1]{%
 \ifx #1\expandafter \@firstoftwo
 \else \expandafter \@secondoftwo
 \fi
}%
\providecommand \natexlab [1]{#1}%
\providecommand \enquote  [1]{``#1''}%
\providecommand \bibnamefont  [1]{#1}%
\providecommand \bibfnamefont [1]{#1}%
\providecommand \citenamefont [1]{#1}%
\providecommand \href@noop [0]{\@secondoftwo}%
\providecommand \href [0]{\begingroup \@sanitize@url \@href}%
\providecommand \@href[1]{\@@startlink{#1}\@@href}%
\providecommand \@@href[1]{\endgroup#1\@@endlink}%
\providecommand \@sanitize@url [0]{\catcode `\\12\catcode `\$12\catcode
  `\&12\catcode `\#12\catcode `\^12\catcode `\_12\catcode `\%12\relax}%
\providecommand \@@startlink[1]{}%
\providecommand \@@endlink[0]{}%
\providecommand \url  [0]{\begingroup\@sanitize@url \@url }%
\providecommand \@url [1]{\endgroup\@href {#1}{\urlprefix }}%
\providecommand \urlprefix  [0]{URL }%
\providecommand \Eprint [0]{\href }%
\providecommand \doibase [0]{https://doi.org/}%
\providecommand \selectlanguage [0]{\@gobble}%
\providecommand \bibinfo  [0]{\@secondoftwo}%
\providecommand \bibfield  [0]{\@secondoftwo}%
\providecommand \translation [1]{[#1]}%
\providecommand \BibitemOpen [0]{}%
\providecommand \bibitemStop [0]{}%
\providecommand \bibitemNoStop [0]{.\EOS\space}%
\providecommand \EOS [0]{\spacefactor3000\relax}%
\providecommand \BibitemShut  [1]{\csname bibitem#1\endcsname}%
\let\auto@bib@innerbib\@empty
\bibitem [{\citenamefont {Aghanim}\ \emph {et~al.}(2018)\citenamefont {Aghanim}
  \emph {et~al.}}]{Aghanim:2018eyx}%
  \BibitemOpen
  \bibfield  {author} {\bibinfo {author} {\bibfnamefont {N.}~\bibnamefont
  {Aghanim}} \emph {et~al.} (\bibinfo {collaboration} {Planck}),\ }\bibfield
  {title} {\bibinfo {title} {{Planck 2018 results. VI. Cosmological
  parameters}},\ }\href@noop {} {\  (\bibinfo {year} {2018})},\ \Eprint
  {https://arxiv.org/abs/1807.06209} {arXiv:1807.06209 [astro-ph.CO]}
  \BibitemShut {NoStop}%
\bibitem [{\citenamefont {Heavens}\ \emph {et~al.}(2017)\citenamefont
  {Heavens}, \citenamefont {Fantaye}, \citenamefont {Sellentin}, \citenamefont
  {Eggers}, \citenamefont {Hosenie}, \citenamefont {Kroon},\ and\ \citenamefont
  {Mootoovaloo}}]{Heavens:2017hkr}%
  \BibitemOpen
  \bibfield  {author} {\bibinfo {author} {\bibfnamefont {A.}~\bibnamefont
  {Heavens}}, \bibinfo {author} {\bibfnamefont {Y.}~\bibnamefont {Fantaye}},
  \bibinfo {author} {\bibfnamefont {E.}~\bibnamefont {Sellentin}}, \bibinfo
  {author} {\bibfnamefont {H.}~\bibnamefont {Eggers}}, \bibinfo {author}
  {\bibfnamefont {Z.}~\bibnamefont {Hosenie}}, \bibinfo {author} {\bibfnamefont
  {S.}~\bibnamefont {Kroon}},\ and\ \bibinfo {author} {\bibfnamefont
  {A.}~\bibnamefont {Mootoovaloo}},\ }\bibfield  {title} {\bibinfo {title} {{No
  evidence for extensions to the standard cosmological model}},\ }\href
  {https://doi.org/10.1103/PhysRevLett.119.101301} {\bibfield  {journal}
  {\bibinfo  {journal} {Phys. Rev. Lett.}\ }\textbf {\bibinfo {volume} {119}},\
  \bibinfo {pages} {101301} (\bibinfo {year} {2017})},\ \Eprint
  {https://arxiv.org/abs/1704.03467} {arXiv:1704.03467 [astro-ph.CO]}
  \BibitemShut {NoStop}%
\bibitem [{\citenamefont {Di~Valentino}\ \emph {et~al.}(2019)\citenamefont
  {Di~Valentino}, \citenamefont {Melchiorri},\ and\ \citenamefont
  {Silk}}]{DiValentino:2019qzk}%
  \BibitemOpen
  \bibfield  {author} {\bibinfo {author} {\bibfnamefont {E.}~\bibnamefont
  {Di~Valentino}}, \bibinfo {author} {\bibfnamefont {A.}~\bibnamefont
  {Melchiorri}},\ and\ \bibinfo {author} {\bibfnamefont {J.}~\bibnamefont
  {Silk}},\ }\bibfield  {title} {\bibinfo {title} {{Planck evidence for a
  closed Universe and a possible crisis for cosmology}},\ }\href
  {https://doi.org/10.1038/s41550-019-0906-9} {\bibfield  {journal} {\bibinfo
  {journal} {Nature Astron.}\ }\textbf {\bibinfo {volume} {4}},\ \bibinfo
  {pages} {196} (\bibinfo {year} {2019})},\ \Eprint
  {https://arxiv.org/abs/1911.02087} {arXiv:1911.02087 [astro-ph.CO]}
  \BibitemShut {NoStop}%
\bibitem [{\citenamefont {Handley}(2021)}]{Handley:2019tkm}%
  \BibitemOpen
  \bibfield  {author} {\bibinfo {author} {\bibfnamefont {W.}~\bibnamefont
  {Handley}},\ }\bibfield  {title} {\bibinfo {title} {{Curvature tension:
  evidence for a closed universe}},\ }\href
  {https://doi.org/10.1103/PhysRevD.103.L041301} {\bibfield  {journal}
  {\bibinfo  {journal} {Phys. Rev. D}\ }\textbf {\bibinfo {volume} {103}},\
  \bibinfo {pages} {L041301} (\bibinfo {year} {2021})},\ \Eprint
  {https://arxiv.org/abs/1908.09139} {arXiv:1908.09139 [astro-ph.CO]}
  \BibitemShut {NoStop}%
\bibitem [{\citenamefont {Di~Valentino}\ \emph
  {et~al.}(2020{\natexlab{a}})\citenamefont {Di~Valentino} \emph
  {et~al.}}]{DiValentino:2020srs}%
  \BibitemOpen
  \bibfield  {author} {\bibinfo {author} {\bibfnamefont {E.}~\bibnamefont
  {Di~Valentino}} \emph {et~al.},\ }\bibfield  {title} {\bibinfo {title}
  {{Cosmology Intertwined IV: The Age of the Universe and its Curvature}},\
  }\href@noop {} {\  (\bibinfo {year} {2020}{\natexlab{a}})},\ \Eprint
  {https://arxiv.org/abs/2008.11286} {arXiv:2008.11286 [astro-ph.CO]}
  \BibitemShut {NoStop}%
\bibitem [{\citenamefont {Virey}\ \emph {et~al.}(2008)\citenamefont {Virey},
  \citenamefont {Talon-Esmieu}, \citenamefont {Ealet}, \citenamefont {Taxil},\
  and\ \citenamefont {Tilquin}}]{Virey:2008nu}%
  \BibitemOpen
  \bibfield  {author} {\bibinfo {author} {\bibfnamefont {J.~M.}\ \bibnamefont
  {Virey}}, \bibinfo {author} {\bibfnamefont {D.}~\bibnamefont {Talon-Esmieu}},
  \bibinfo {author} {\bibfnamefont {A.}~\bibnamefont {Ealet}}, \bibinfo
  {author} {\bibfnamefont {P.}~\bibnamefont {Taxil}},\ and\ \bibinfo {author}
  {\bibfnamefont {A.}~\bibnamefont {Tilquin}},\ }\bibfield  {title} {\bibinfo
  {title} {{On the determination of curvature and dynamical Dark Energy}},\
  }\href {https://doi.org/10.1088/1475-7516/2008/12/008} {\bibfield  {journal}
  {\bibinfo  {journal} {JCAP}\ }\textbf {\bibinfo {volume} {12}},\ \bibinfo
  {pages} {008}},\ \Eprint {https://arxiv.org/abs/0802.4407} {arXiv:0802.4407
  [astro-ph]} \BibitemShut {NoStop}%
\bibitem [{\citenamefont {Clarkson}\ \emph {et~al.}(2008)\citenamefont
  {Clarkson}, \citenamefont {Bassett},\ and\ \citenamefont
  {Lu}}]{Clarkson:2007pz}%
  \BibitemOpen
  \bibfield  {author} {\bibinfo {author} {\bibfnamefont {C.}~\bibnamefont
  {Clarkson}}, \bibinfo {author} {\bibfnamefont {B.}~\bibnamefont {Bassett}},\
  and\ \bibinfo {author} {\bibfnamefont {T.~H.-C.}\ \bibnamefont {Lu}},\
  }\bibfield  {title} {\bibinfo {title} {{A general test of the Copernican
  Principle}},\ }\href {https://doi.org/10.1103/PhysRevLett.101.011301}
  {\bibfield  {journal} {\bibinfo  {journal} {Phys. Rev. Lett.}\ }\textbf
  {\bibinfo {volume} {101}},\ \bibinfo {pages} {011301} (\bibinfo {year}
  {2008})},\ \Eprint {https://arxiv.org/abs/0712.3457} {arXiv:0712.3457
  [astro-ph]} \BibitemShut {NoStop}%
\bibitem [{\citenamefont {Shafieloo}\ and\ \citenamefont
  {Clarkson}(2010)}]{Shafieloo:2009hi}%
  \BibitemOpen
  \bibfield  {author} {\bibinfo {author} {\bibfnamefont {A.}~\bibnamefont
  {Shafieloo}}\ and\ \bibinfo {author} {\bibfnamefont {C.}~\bibnamefont
  {Clarkson}},\ }\bibfield  {title} {\bibinfo {title} {{Model independent tests
  of the standard cosmological model}},\ }\href
  {https://doi.org/10.1103/PhysRevD.81.083537} {\bibfield  {journal} {\bibinfo
  {journal} {Phys. Rev. D}\ }\textbf {\bibinfo {volume} {81}},\ \bibinfo
  {pages} {083537} (\bibinfo {year} {2010})},\ \Eprint
  {https://arxiv.org/abs/0911.4858} {arXiv:0911.4858 [astro-ph.CO]}
  \BibitemShut {NoStop}%
\bibitem [{\citenamefont {Mortsell}\ and\ \citenamefont
  {Jonsson}(2011)}]{Mortsell:2011yk}%
  \BibitemOpen
  \bibfield  {author} {\bibinfo {author} {\bibfnamefont {E.}~\bibnamefont
  {Mortsell}}\ and\ \bibinfo {author} {\bibfnamefont {J.}~\bibnamefont
  {Jonsson}},\ }\bibfield  {title} {\bibinfo {title} {{A model independent
  measure of the large scale curvature of the Universe}},\ }\href@noop {} {\
  (\bibinfo {year} {2011})},\ \Eprint {https://arxiv.org/abs/1102.4485}
  {arXiv:1102.4485 [astro-ph.CO]} \BibitemShut {NoStop}%
\bibitem [{\citenamefont {Sapone}\ \emph {et~al.}(2014)\citenamefont {Sapone},
  \citenamefont {Majerotto},\ and\ \citenamefont {Nesseris}}]{Sapone:2014nna}%
  \BibitemOpen
  \bibfield  {author} {\bibinfo {author} {\bibfnamefont {D.}~\bibnamefont
  {Sapone}}, \bibinfo {author} {\bibfnamefont {E.}~\bibnamefont {Majerotto}},\
  and\ \bibinfo {author} {\bibfnamefont {S.}~\bibnamefont {Nesseris}},\
  }\bibfield  {title} {\bibinfo {title} {{Curvature versus distances: Testing
  the FLRW cosmology}},\ }\href {https://doi.org/10.1103/PhysRevD.90.023012}
  {\bibfield  {journal} {\bibinfo  {journal} {Phys. Rev. D}\ }\textbf {\bibinfo
  {volume} {90}},\ \bibinfo {pages} {023012} (\bibinfo {year} {2014})},\
  \Eprint {https://arxiv.org/abs/1402.2236} {arXiv:1402.2236 [astro-ph.CO]}
  \BibitemShut {NoStop}%
\bibitem [{\citenamefont {R\"as\"anen}\ \emph {et~al.}(2015)\citenamefont
  {R\"as\"anen}, \citenamefont {Bolejko},\ and\ \citenamefont
  {Finoguenov}}]{Rasanen:2014mca}%
  \BibitemOpen
  \bibfield  {author} {\bibinfo {author} {\bibfnamefont {S.}~\bibnamefont
  {R\"as\"anen}}, \bibinfo {author} {\bibfnamefont {K.}~\bibnamefont
  {Bolejko}},\ and\ \bibinfo {author} {\bibfnamefont {A.}~\bibnamefont
  {Finoguenov}},\ }\bibfield  {title} {\bibinfo {title} {{New Test of the
  Friedmann-Lema\^\i{}tre-Robertson-Walker Metric Using the Distance Sum
  Rule}},\ }\href {https://doi.org/10.1103/PhysRevLett.115.101301} {\bibfield
  {journal} {\bibinfo  {journal} {Phys. Rev. Lett.}\ }\textbf {\bibinfo
  {volume} {115}},\ \bibinfo {pages} {101301} (\bibinfo {year} {2015})},\
  \Eprint {https://arxiv.org/abs/1412.4976} {arXiv:1412.4976 [astro-ph.CO]}
  \BibitemShut {NoStop}%
\bibitem [{\citenamefont {L'Huillier}\ and\ \citenamefont
  {Shafieloo}(2017)}]{LHuillier:2016mtc}%
  \BibitemOpen
  \bibfield  {author} {\bibinfo {author} {\bibfnamefont {B.}~\bibnamefont
  {L'Huillier}}\ and\ \bibinfo {author} {\bibfnamefont {A.}~\bibnamefont
  {Shafieloo}},\ }\bibfield  {title} {\bibinfo {title} {{Model-independent test
  of the FLRW metric, the flatness of the Universe, and non-local measurement
  of $H_0r_\mathrm{d}$}},\ }\href
  {https://doi.org/10.1088/1475-7516/2017/01/015} {\bibfield  {journal}
  {\bibinfo  {journal} {JCAP}\ }\textbf {\bibinfo {volume} {01}},\ \bibinfo
  {pages} {015}},\ \Eprint {https://arxiv.org/abs/1606.06832} {arXiv:1606.06832
  [astro-ph.CO]} \BibitemShut {NoStop}%
\bibitem [{\citenamefont {Denissenya}\ \emph {et~al.}(2018)\citenamefont
  {Denissenya}, \citenamefont {Linder},\ and\ \citenamefont
  {Shafieloo}}]{Denissenya:2018zcv}%
  \BibitemOpen
  \bibfield  {author} {\bibinfo {author} {\bibfnamefont {M.}~\bibnamefont
  {Denissenya}}, \bibinfo {author} {\bibfnamefont {E.~V.}\ \bibnamefont
  {Linder}},\ and\ \bibinfo {author} {\bibfnamefont {A.}~\bibnamefont
  {Shafieloo}},\ }\bibfield  {title} {\bibinfo {title} {{Cosmic Curvature
  Tested Directly from Observations}},\ }\href
  {https://doi.org/10.1088/1475-7516/2018/03/041} {\bibfield  {journal}
  {\bibinfo  {journal} {JCAP}\ }\textbf {\bibinfo {volume} {03}},\ \bibinfo
  {pages} {041}},\ \Eprint {https://arxiv.org/abs/1802.04816} {arXiv:1802.04816
  [astro-ph.CO]} \BibitemShut {NoStop}%
\bibitem [{\citenamefont {Park}\ and\ \citenamefont
  {Ratra}(2019)}]{Park:2017xbl}%
  \BibitemOpen
  \bibfield  {author} {\bibinfo {author} {\bibfnamefont {C.-G.}\ \bibnamefont
  {Park}}\ and\ \bibinfo {author} {\bibfnamefont {B.}~\bibnamefont {Ratra}},\
  }\bibfield  {title} {\bibinfo {title} {{Using the tilted flat-$\Lambda$CDM
  and the untilted non-flat $\Lambda$CDM inflation models to measure
  cosmological parameters from a compilation of observational data}},\ }\href
  {https://doi.org/10.3847/1538-4357/ab3641} {\bibfield  {journal} {\bibinfo
  {journal} {Astrophys. J.}\ }\textbf {\bibinfo {volume} {882}},\ \bibinfo
  {pages} {158} (\bibinfo {year} {2019})},\ \Eprint
  {https://arxiv.org/abs/1801.00213} {arXiv:1801.00213 [astro-ph.CO]}
  \BibitemShut {NoStop}%
\bibitem [{\citenamefont {Cao}\ \emph {et~al.}(2021)\citenamefont {Cao},
  \citenamefont {Ryan},\ and\ \citenamefont {Ratra}}]{Cao:2021ldv}%
  \BibitemOpen
  \bibfield  {author} {\bibinfo {author} {\bibfnamefont {S.}~\bibnamefont
  {Cao}}, \bibinfo {author} {\bibfnamefont {J.}~\bibnamefont {Ryan}},\ and\
  \bibinfo {author} {\bibfnamefont {B.}~\bibnamefont {Ratra}},\ }\bibfield
  {title} {\bibinfo {title} {{Using Pantheon and DES supernova, baryon acoustic
  oscillation, and Hubble parameter data to constrain the Hubble constant, dark
  energy dynamics, and spatial curvature}},\ }\href@noop {} {\  (\bibinfo
  {year} {2021})},\ \Eprint {https://arxiv.org/abs/2101.08817}
  {arXiv:2101.08817 [astro-ph.CO]} \BibitemShut {NoStop}%
\bibitem [{\citenamefont {Khadka}\ and\ \citenamefont
  {Ratra}(2020)}]{Khadka:2020tlm}%
  \BibitemOpen
  \bibfield  {author} {\bibinfo {author} {\bibfnamefont {N.}~\bibnamefont
  {Khadka}}\ and\ \bibinfo {author} {\bibfnamefont {B.}~\bibnamefont {Ratra}},\
  }\bibfield  {title} {\bibinfo {title} {{Determining the range of validity of
  quasar X-ray and UV flux measurements for constraining cosmological model
  parameters}},\ }\href@noop {} {\  (\bibinfo {year} {2020})},\ \Eprint
  {https://arxiv.org/abs/2012.09291} {arXiv:2012.09291 [astro-ph.CO]}
  \BibitemShut {NoStop}%
\bibitem [{\citenamefont {Guth}\ \emph {et~al.}(2014)\citenamefont {Guth},
  \citenamefont {Kaiser},\ and\ \citenamefont {Nomura}}]{Guth:2013sya}%
  \BibitemOpen
  \bibfield  {author} {\bibinfo {author} {\bibfnamefont {A.~H.}\ \bibnamefont
  {Guth}}, \bibinfo {author} {\bibfnamefont {D.~I.}\ \bibnamefont {Kaiser}},\
  and\ \bibinfo {author} {\bibfnamefont {Y.}~\bibnamefont {Nomura}},\
  }\bibfield  {title} {\bibinfo {title} {{Inflationary paradigm after Planck
  2013}},\ }\href {https://doi.org/10.1016/j.physletb.2014.03.020} {\bibfield
  {journal} {\bibinfo  {journal} {Phys. Lett. B}\ }\textbf {\bibinfo {volume}
  {733}},\ \bibinfo {pages} {112} (\bibinfo {year} {2014})},\ \Eprint
  {https://arxiv.org/abs/1312.7619} {arXiv:1312.7619 [astro-ph.CO]}
  \BibitemShut {NoStop}%
\bibitem [{\citenamefont {Di~Dio}\ \emph {et~al.}(2016)\citenamefont {Di~Dio},
  \citenamefont {Montanari}, \citenamefont {Raccanelli}, \citenamefont
  {Durrer}, \citenamefont {Kamionkowski},\ and\ \citenamefont
  {Lesgourgues}}]{DiDio:2016ykq}%
  \BibitemOpen
  \bibfield  {author} {\bibinfo {author} {\bibfnamefont {E.}~\bibnamefont
  {Di~Dio}}, \bibinfo {author} {\bibfnamefont {F.}~\bibnamefont {Montanari}},
  \bibinfo {author} {\bibfnamefont {A.}~\bibnamefont {Raccanelli}}, \bibinfo
  {author} {\bibfnamefont {R.}~\bibnamefont {Durrer}}, \bibinfo {author}
  {\bibfnamefont {M.}~\bibnamefont {Kamionkowski}},\ and\ \bibinfo {author}
  {\bibfnamefont {J.}~\bibnamefont {Lesgourgues}},\ }\bibfield  {title}
  {\bibinfo {title} {{Curvature constraints from Large Scale Structure}},\
  }\href {https://doi.org/10.1088/1475-7516/2016/06/013} {\bibfield  {journal}
  {\bibinfo  {journal} {JCAP}\ }\textbf {\bibinfo {volume} {06}},\ \bibinfo
  {pages} {013}},\ \Eprint {https://arxiv.org/abs/1603.09073} {arXiv:1603.09073
  [astro-ph.CO]} \BibitemShut {NoStop}%
\bibitem [{\citenamefont {Vardanyan}\ \emph {et~al.}(2009)\citenamefont
  {Vardanyan}, \citenamefont {Trotta},\ and\ \citenamefont
  {Silk}}]{Vardanyan:2009ft}%
  \BibitemOpen
  \bibfield  {author} {\bibinfo {author} {\bibfnamefont {M.}~\bibnamefont
  {Vardanyan}}, \bibinfo {author} {\bibfnamefont {R.}~\bibnamefont {Trotta}},\
  and\ \bibinfo {author} {\bibfnamefont {J.}~\bibnamefont {Silk}},\ }\bibfield
  {title} {\bibinfo {title} {{How flat can you get? A model comparison
  perspective on the curvature of the Universe}},\ }\href
  {https://doi.org/10.1111/j.1365-2966.2009.14938.x} {\bibfield  {journal}
  {\bibinfo  {journal} {Mon. Not. Roy. Astron. Soc.}\ }\textbf {\bibinfo
  {volume} {397}},\ \bibinfo {pages} {431} (\bibinfo {year} {2009})},\ \Eprint
  {https://arxiv.org/abs/0901.3354} {arXiv:0901.3354 [astro-ph.CO]}
  \BibitemShut {NoStop}%
\bibitem [{\citenamefont {Bertone}\ and\ \citenamefont
  {Hooper}(2018)}]{Bertone:2016nfn}%
  \BibitemOpen
  \bibfield  {author} {\bibinfo {author} {\bibfnamefont {G.}~\bibnamefont
  {Bertone}}\ and\ \bibinfo {author} {\bibfnamefont {D.}~\bibnamefont
  {Hooper}},\ }\bibfield  {title} {\bibinfo {title} {{History of dark
  matter}},\ }\href {https://doi.org/10.1103/RevModPhys.90.045002} {\bibfield
  {journal} {\bibinfo  {journal} {Rev. Mod. Phys.}\ }\textbf {\bibinfo {volume}
  {90}},\ \bibinfo {pages} {045002} (\bibinfo {year} {2018})},\ \Eprint
  {https://arxiv.org/abs/1605.04909} {arXiv:1605.04909 [astro-ph.CO]}
  \BibitemShut {NoStop}%
\bibitem [{\citenamefont {Weinberg}(1989)}]{Weinberg:1988cp}%
  \BibitemOpen
  \bibfield  {author} {\bibinfo {author} {\bibfnamefont {S.}~\bibnamefont
  {Weinberg}},\ }\bibfield  {title} {\bibinfo {title} {{The Cosmological
  Constant Problem}},\ }\href {https://doi.org/10.1103/RevModPhys.61.1}
  {\bibfield  {journal} {\bibinfo  {journal} {Rev. Mod. Phys.}\ }\textbf
  {\bibinfo {volume} {61}},\ \bibinfo {pages} {1} (\bibinfo {year}
  {1989})}\BibitemShut {NoStop}%
\bibitem [{\citenamefont {Carroll}(2001)}]{Carroll:2000fy}%
  \BibitemOpen
  \bibfield  {author} {\bibinfo {author} {\bibfnamefont {S.~M.}\ \bibnamefont
  {Carroll}},\ }\bibfield  {title} {\bibinfo {title} {{The Cosmological
  constant}},\ }\href {https://doi.org/10.12942/lrr-2001-1} {\bibfield
  {journal} {\bibinfo  {journal} {Living Rev. Rel.}\ }\textbf {\bibinfo
  {volume} {4}},\ \bibinfo {pages} {1} (\bibinfo {year} {2001})},\ \Eprint
  {https://arxiv.org/abs/astro-ph/0004075} {arXiv:astro-ph/0004075}
  \BibitemShut {NoStop}%
\bibitem [{\citenamefont {Di~Valentino}\ \emph
  {et~al.}(2020{\natexlab{b}})\citenamefont {Di~Valentino} \emph
  {et~al.}}]{DiValentino:2020vhf}%
  \BibitemOpen
  \bibfield  {author} {\bibinfo {author} {\bibfnamefont {E.}~\bibnamefont
  {Di~Valentino}} \emph {et~al.},\ }\bibfield  {title} {\bibinfo {title}
  {{Cosmology Intertwined I: Perspectives for the Next Decade}},\ }\href@noop
  {} {\  (\bibinfo {year} {2020}{\natexlab{b}})},\ \Eprint
  {https://arxiv.org/abs/2008.11283} {arXiv:2008.11283 [astro-ph.CO]}
  \BibitemShut {NoStop}%
\bibitem [{\citenamefont {Gubitosi}\ \emph {et~al.}(2013)\citenamefont
  {Gubitosi}, \citenamefont {Piazza},\ and\ \citenamefont
  {Vernizzi}}]{Gubitosi:2012hu}%
  \BibitemOpen
  \bibfield  {author} {\bibinfo {author} {\bibfnamefont {G.}~\bibnamefont
  {Gubitosi}}, \bibinfo {author} {\bibfnamefont {F.}~\bibnamefont {Piazza}},\
  and\ \bibinfo {author} {\bibfnamefont {F.}~\bibnamefont {Vernizzi}},\
  }\bibfield  {title} {\bibinfo {title} {{The Effective Field Theory of Dark
  Energy}},\ }\href {https://doi.org/10.1088/1475-7516/2013/02/032} {\bibfield
  {journal} {\bibinfo  {journal} {JCAP}\ }\textbf {\bibinfo {volume} {02}},\
  \bibinfo {pages} {032}},\ \Eprint {https://arxiv.org/abs/1210.0201}
  {arXiv:1210.0201 [hep-th]} \BibitemShut {NoStop}%
\bibitem [{\citenamefont {Hu}\ \emph {et~al.}(2014)\citenamefont {Hu},
  \citenamefont {Raveri}, \citenamefont {Frusciante},\ and\ \citenamefont
  {Silvestri}}]{Hu:2013twa}%
  \BibitemOpen
  \bibfield  {author} {\bibinfo {author} {\bibfnamefont {B.}~\bibnamefont
  {Hu}}, \bibinfo {author} {\bibfnamefont {M.}~\bibnamefont {Raveri}}, \bibinfo
  {author} {\bibfnamefont {N.}~\bibnamefont {Frusciante}},\ and\ \bibinfo
  {author} {\bibfnamefont {A.}~\bibnamefont {Silvestri}},\ }\bibfield  {title}
  {\bibinfo {title} {{Effective Field Theory of Cosmic Acceleration: an
  implementation in CAMB}},\ }\href
  {https://doi.org/10.1103/PhysRevD.89.103530} {\bibfield  {journal} {\bibinfo
  {journal} {Phys. Rev. D}\ }\textbf {\bibinfo {volume} {89}},\ \bibinfo
  {pages} {103530} (\bibinfo {year} {2014})},\ \Eprint
  {https://arxiv.org/abs/1312.5742} {arXiv:1312.5742 [astro-ph.CO]}
  \BibitemShut {NoStop}%
\bibitem [{\citenamefont {Arjona}\ \emph
  {et~al.}(2019{\natexlab{a}})\citenamefont {Arjona}, \citenamefont {Cardona},\
  and\ \citenamefont {Nesseris}}]{Arjona:2018jhh}%
  \BibitemOpen
  \bibfield  {author} {\bibinfo {author} {\bibfnamefont {R.}~\bibnamefont
  {Arjona}}, \bibinfo {author} {\bibfnamefont {W.}~\bibnamefont {Cardona}},\
  and\ \bibinfo {author} {\bibfnamefont {S.}~\bibnamefont {Nesseris}},\
  }\bibfield  {title} {\bibinfo {title} {{Unraveling the effective fluid
  approach for $f(R)$ models in the subhorizon approximation}},\ }\href
  {https://doi.org/10.1103/PhysRevD.99.043516} {\bibfield  {journal} {\bibinfo
  {journal} {Phys. Rev. D}\ }\textbf {\bibinfo {volume} {99}},\ \bibinfo
  {pages} {043516} (\bibinfo {year} {2019}{\natexlab{a}})},\ \Eprint
  {https://arxiv.org/abs/1811.02469} {arXiv:1811.02469 [astro-ph.CO]}
  \BibitemShut {NoStop}%
\bibitem [{\citenamefont {Arjona}\ \emph
  {et~al.}(2019{\natexlab{b}})\citenamefont {Arjona}, \citenamefont {Cardona},\
  and\ \citenamefont {Nesseris}}]{Arjona:2019rfn}%
  \BibitemOpen
  \bibfield  {author} {\bibinfo {author} {\bibfnamefont {R.}~\bibnamefont
  {Arjona}}, \bibinfo {author} {\bibfnamefont {W.}~\bibnamefont {Cardona}},\
  and\ \bibinfo {author} {\bibfnamefont {S.}~\bibnamefont {Nesseris}},\
  }\bibfield  {title} {\bibinfo {title} {{Designing Horndeski and the effective
  fluid approach}},\ }\href {https://doi.org/10.1103/PhysRevD.100.063526}
  {\bibfield  {journal} {\bibinfo  {journal} {Phys. Rev. D}\ }\textbf {\bibinfo
  {volume} {100}},\ \bibinfo {pages} {063526} (\bibinfo {year}
  {2019}{\natexlab{b}})},\ \Eprint {https://arxiv.org/abs/1904.06294}
  {arXiv:1904.06294 [astro-ph.CO]} \BibitemShut {NoStop}%
\bibitem [{\citenamefont {Arjona}(2020{\natexlab{a}})}]{Arjona:2020gtm}%
  \BibitemOpen
  \bibfield  {author} {\bibinfo {author} {\bibfnamefont {R.}~\bibnamefont
  {Arjona}},\ }\bibfield  {title} {\bibinfo {title} {{The effective fluid
  approach for modified gravity}},\ }in\ \href@noop {} {\emph {\bibinfo
  {booktitle} {{2nd Hermann Minkowski Meeting on the Foundations of Spacetime
  Physics}}}}\ (\bibinfo {year} {2020})\ \Eprint
  {https://arxiv.org/abs/2010.04764} {arXiv:2010.04764 [astro-ph.CO]}
  \BibitemShut {NoStop}%
\bibitem [{\citenamefont {Cardona}\ \emph {et~al.}(2020)\citenamefont
  {Cardona}, \citenamefont {Arjona}, \citenamefont {Estrada},\ and\
  \citenamefont {Nesseris}}]{Cardona:2020ama}%
  \BibitemOpen
  \bibfield  {author} {\bibinfo {author} {\bibfnamefont {W.}~\bibnamefont
  {Cardona}}, \bibinfo {author} {\bibfnamefont {R.}~\bibnamefont {Arjona}},
  \bibinfo {author} {\bibfnamefont {A.}~\bibnamefont {Estrada}},\ and\ \bibinfo
  {author} {\bibfnamefont {S.}~\bibnamefont {Nesseris}},\ }\bibfield  {title}
  {\bibinfo {title} {{Cosmological constraints with the Effective Fluid
  approach for Modified Gravity}},\ }\href@noop {} {\  (\bibinfo {year}
  {2020})},\ \Eprint {https://arxiv.org/abs/2012.05282} {arXiv:2012.05282
  [astro-ph.CO]} \BibitemShut {NoStop}%
\bibitem [{\citenamefont {Nesseris}\ and\ \citenamefont
  {Shafieloo}(2010)}]{Nesseris:2010ep}%
  \BibitemOpen
  \bibfield  {author} {\bibinfo {author} {\bibfnamefont {S.}~\bibnamefont
  {Nesseris}}\ and\ \bibinfo {author} {\bibfnamefont {A.}~\bibnamefont
  {Shafieloo}},\ }\bibfield  {title} {\bibinfo {title} {{A model independent
  null test on the cosmological constant}},\ }\href
  {https://doi.org/10.1111/j.1365-2966.2010.17254.x} {\bibfield  {journal}
  {\bibinfo  {journal} {Mon. Not. Roy. Astron. Soc.}\ }\textbf {\bibinfo
  {volume} {408}},\ \bibinfo {pages} {1879} (\bibinfo {year} {2010})},\ \Eprint
  {https://arxiv.org/abs/1004.0960} {arXiv:1004.0960 [astro-ph.CO]}
  \BibitemShut {NoStop}%
\bibitem [{\citenamefont {Ntampaka}\ \emph {et~al.}(2019)\citenamefont
  {Ntampaka} \emph {et~al.}}]{Ntampaka:2019udw}%
  \BibitemOpen
  \bibfield  {author} {\bibinfo {author} {\bibfnamefont {M.}~\bibnamefont
  {Ntampaka}} \emph {et~al.},\ }\bibfield  {title} {\bibinfo {title} {{The Role
  of Machine Learning in the Next Decade of Cosmology}},\ }\href@noop {} {\
  (\bibinfo {year} {2019})},\ \Eprint {https://arxiv.org/abs/1902.10159}
  {arXiv:1902.10159 [astro-ph.IM]} \BibitemShut {NoStop}%
\bibitem [{\citenamefont {Marra}\ and\ \citenamefont
  {Sapone}(2018)}]{Marra:2017pst}%
  \BibitemOpen
  \bibfield  {author} {\bibinfo {author} {\bibfnamefont {V.}~\bibnamefont
  {Marra}}\ and\ \bibinfo {author} {\bibfnamefont {D.}~\bibnamefont {Sapone}},\
  }\bibfield  {title} {\bibinfo {title} {{Null tests of the standard model
  using the linear model formalism}},\ }\href
  {https://doi.org/10.1103/PhysRevD.97.083510} {\bibfield  {journal} {\bibinfo
  {journal} {Phys. Rev. D}\ }\textbf {\bibinfo {volume} {97}},\ \bibinfo
  {pages} {083510} (\bibinfo {year} {2018})},\ \Eprint
  {https://arxiv.org/abs/1712.09676} {arXiv:1712.09676 [astro-ph.CO]}
  \BibitemShut {NoStop}%
\bibitem [{\citenamefont {Sahni}\ \emph {et~al.}(2008)\citenamefont {Sahni},
  \citenamefont {Shafieloo},\ and\ \citenamefont {Starobinsky}}]{Sahni:2008xx}%
  \BibitemOpen
  \bibfield  {author} {\bibinfo {author} {\bibfnamefont {V.}~\bibnamefont
  {Sahni}}, \bibinfo {author} {\bibfnamefont {A.}~\bibnamefont {Shafieloo}},\
  and\ \bibinfo {author} {\bibfnamefont {A.~A.}\ \bibnamefont {Starobinsky}},\
  }\bibfield  {title} {\bibinfo {title} {{Two new diagnostics of dark
  energy}},\ }\href {https://doi.org/10.1103/PhysRevD.78.103502} {\bibfield
  {journal} {\bibinfo  {journal} {Phys. Rev.}\ }\textbf {\bibinfo {volume}
  {D78}},\ \bibinfo {pages} {103502} (\bibinfo {year} {2008})},\ \Eprint
  {https://arxiv.org/abs/0807.3548} {arXiv:0807.3548 [astro-ph]} \BibitemShut
  {NoStop}%
\bibitem [{\citenamefont {Zunckel}\ and\ \citenamefont
  {Clarkson}(2008)}]{Zunckel:2008ti}%
  \BibitemOpen
  \bibfield  {author} {\bibinfo {author} {\bibfnamefont {C.}~\bibnamefont
  {Zunckel}}\ and\ \bibinfo {author} {\bibfnamefont {C.}~\bibnamefont
  {Clarkson}},\ }\bibfield  {title} {\bibinfo {title} {{Consistency Tests for
  the Cosmological Constant}},\ }\href
  {https://doi.org/10.1103/PhysRevLett.101.181301} {\bibfield  {journal}
  {\bibinfo  {journal} {Phys. Rev. Lett.}\ }\textbf {\bibinfo {volume} {101}},\
  \bibinfo {pages} {181301} (\bibinfo {year} {2008})},\ \Eprint
  {https://arxiv.org/abs/0807.4304} {arXiv:0807.4304 [astro-ph]} \BibitemShut
  {NoStop}%
\bibitem [{\citenamefont {von Marttens}\ \emph {et~al.}(2019)\citenamefont {von
  Marttens}, \citenamefont {Marra}, \citenamefont {Casarini}, \citenamefont
  {Gonzalez},\ and\ \citenamefont {Alcaniz}}]{vonMarttens:2018bvz}%
  \BibitemOpen
  \bibfield  {author} {\bibinfo {author} {\bibfnamefont {R.}~\bibnamefont {von
  Marttens}}, \bibinfo {author} {\bibfnamefont {V.}~\bibnamefont {Marra}},
  \bibinfo {author} {\bibfnamefont {L.}~\bibnamefont {Casarini}}, \bibinfo
  {author} {\bibfnamefont {J.~E.}\ \bibnamefont {Gonzalez}},\ and\ \bibinfo
  {author} {\bibfnamefont {J.}~\bibnamefont {Alcaniz}},\ }\bibfield  {title}
  {\bibinfo {title} {{Null test for interactions in the dark sector}},\ }\href
  {https://doi.org/10.1103/PhysRevD.99.043521} {\bibfield  {journal} {\bibinfo
  {journal} {Phys. Rev.}\ }\textbf {\bibinfo {volume} {D99}},\ \bibinfo {pages}
  {043521} (\bibinfo {year} {2019})},\ \Eprint
  {https://arxiv.org/abs/1812.02333} {arXiv:1812.02333 [astro-ph.CO]}
  \BibitemShut {NoStop}%
\bibitem [{\citenamefont {Nesseris}\ and\ \citenamefont
  {Sapone}(2015)}]{Nesseris:2014mfa}%
  \BibitemOpen
  \bibfield  {author} {\bibinfo {author} {\bibfnamefont {S.}~\bibnamefont
  {Nesseris}}\ and\ \bibinfo {author} {\bibfnamefont {D.}~\bibnamefont
  {Sapone}},\ }\bibfield  {title} {\bibinfo {title} {{Novel null-test for the
  $\Lambda$ cold dark matter model with growth-rate data}},\ }\href
  {https://doi.org/10.1142/S0218271815500455} {\bibfield  {journal} {\bibinfo
  {journal} {Int. J. Mod. Phys.}\ }\textbf {\bibinfo {volume} {D24}},\ \bibinfo
  {pages} {1550045} (\bibinfo {year} {2015})},\ \Eprint
  {https://arxiv.org/abs/1409.3697} {arXiv:1409.3697 [astro-ph.CO]}
  \BibitemShut {NoStop}%
\bibitem [{\citenamefont {Benisty}(2021)}]{Benisty:2020kdt}%
  \BibitemOpen
  \bibfield  {author} {\bibinfo {author} {\bibfnamefont {D.}~\bibnamefont
  {Benisty}},\ }\bibfield  {title} {\bibinfo {title} {{Quantifying the $S_8$
  tension with the Redshift Space Distortion data set}},\ }\href
  {https://doi.org/10.1016/j.dark.2020.100766} {\bibfield  {journal} {\bibinfo
  {journal} {Phys. Dark Univ.}\ }\textbf {\bibinfo {volume} {31}},\ \bibinfo
  {pages} {100766} (\bibinfo {year} {2021})},\ \Eprint
  {https://arxiv.org/abs/2005.03751} {arXiv:2005.03751 [astro-ph.CO]}
  \BibitemShut {NoStop}%
\bibitem [{\citenamefont {Yahya}\ \emph {et~al.}(2014)\citenamefont {Yahya},
  \citenamefont {Seikel}, \citenamefont {Clarkson}, \citenamefont {Maartens},\
  and\ \citenamefont {Smith}}]{Yahya:2013xma}%
  \BibitemOpen
  \bibfield  {author} {\bibinfo {author} {\bibfnamefont {S.}~\bibnamefont
  {Yahya}}, \bibinfo {author} {\bibfnamefont {M.}~\bibnamefont {Seikel}},
  \bibinfo {author} {\bibfnamefont {C.}~\bibnamefont {Clarkson}}, \bibinfo
  {author} {\bibfnamefont {R.}~\bibnamefont {Maartens}},\ and\ \bibinfo
  {author} {\bibfnamefont {M.}~\bibnamefont {Smith}},\ }\bibfield  {title}
  {\bibinfo {title} {{Null tests of the cosmological constant using
  supernovae}},\ }\href {https://doi.org/10.1103/PhysRevD.89.023503} {\bibfield
   {journal} {\bibinfo  {journal} {Phys. Rev.}\ }\textbf {\bibinfo {volume}
  {D89}},\ \bibinfo {pages} {023503} (\bibinfo {year} {2014})},\ \Eprint
  {https://arxiv.org/abs/1308.4099} {arXiv:1308.4099 [astro-ph.CO]}
  \BibitemShut {NoStop}%
\bibitem [{\citenamefont {Cai}\ \emph {et~al.}(2016)\citenamefont {Cai},
  \citenamefont {Guo},\ and\ \citenamefont {Yang}}]{Cai:2015pia}%
  \BibitemOpen
  \bibfield  {author} {\bibinfo {author} {\bibfnamefont {R.-G.}\ \bibnamefont
  {Cai}}, \bibinfo {author} {\bibfnamefont {Z.-K.}\ \bibnamefont {Guo}},\ and\
  \bibinfo {author} {\bibfnamefont {T.}~\bibnamefont {Yang}},\ }\bibfield
  {title} {\bibinfo {title} {{Null test of the cosmic curvature using $H(z)$
  and supernovae data}},\ }\href {https://doi.org/10.1103/PhysRevD.93.043517}
  {\bibfield  {journal} {\bibinfo  {journal} {Phys. Rev.}\ }\textbf {\bibinfo
  {volume} {D93}},\ \bibinfo {pages} {043517} (\bibinfo {year} {2016})},\
  \Eprint {https://arxiv.org/abs/1509.06283} {arXiv:1509.06283 [astro-ph.CO]}
  \BibitemShut {NoStop}%
\bibitem [{\citenamefont {Benisty}\ and\ \citenamefont
  {Staicova}(2021)}]{Benisty:2020otr}%
  \BibitemOpen
  \bibfield  {author} {\bibinfo {author} {\bibfnamefont {D.}~\bibnamefont
  {Benisty}}\ and\ \bibinfo {author} {\bibfnamefont {D.}~\bibnamefont
  {Staicova}},\ }\bibfield  {title} {\bibinfo {title} {{Testing Late Time
  Cosmic Acceleration with uncorrelated Baryon Acoustic Oscillations
  dataset}},\ }\href {https://doi.org/10.1051/0004-6361/202039502} {\bibfield
  {journal} {\bibinfo  {journal} {Astron. Astrophys.}\ }\textbf {\bibinfo
  {volume} {647}},\ \bibinfo {pages} {A38} (\bibinfo {year} {2021})},\ \Eprint
  {https://arxiv.org/abs/2009.10701} {arXiv:2009.10701 [astro-ph.CO]}
  \BibitemShut {NoStop}%
\bibitem [{\citenamefont {Li}\ \emph {et~al.}(2014)\citenamefont {Li},
  \citenamefont {Li}, \citenamefont {Zhang},\ and\ \citenamefont
  {Li}}]{Li:2014yza}%
  \BibitemOpen
  \bibfield  {author} {\bibinfo {author} {\bibfnamefont {Y.-L.}\ \bibnamefont
  {Li}}, \bibinfo {author} {\bibfnamefont {S.-Y.}\ \bibnamefont {Li}}, \bibinfo
  {author} {\bibfnamefont {T.-J.}\ \bibnamefont {Zhang}},\ and\ \bibinfo
  {author} {\bibfnamefont {T.-P.}\ \bibnamefont {Li}},\ }\bibfield  {title}
  {\bibinfo {title} {{Model-independent determination of curvature parameter by
  using $H(z)$ and $D_A(z)$ data pairs from BAO measurements}},\ }\href
  {https://doi.org/10.1088/2041-8205/789/1/L15} {\bibfield  {journal} {\bibinfo
   {journal} {Astrophys. J. Lett.}\ }\textbf {\bibinfo {volume} {789}},\
  \bibinfo {pages} {L15} (\bibinfo {year} {2014})},\ \Eprint
  {https://arxiv.org/abs/1404.0773} {arXiv:1404.0773 [astro-ph.CO]}
  \BibitemShut {NoStop}%
\bibitem [{\citenamefont {Franco}\ \emph {et~al.}(2019)\citenamefont {Franco},
  \citenamefont {Bonvin},\ and\ \citenamefont {Clarkson}}]{Franco:2019wbj}%
  \BibitemOpen
  \bibfield  {author} {\bibinfo {author} {\bibfnamefont {F.~O.}\ \bibnamefont
  {Franco}}, \bibinfo {author} {\bibfnamefont {C.}~\bibnamefont {Bonvin}},\
  and\ \bibinfo {author} {\bibfnamefont {C.}~\bibnamefont {Clarkson}},\
  }\bibfield  {title} {\bibinfo {title} {{A null test to probe the
  scale-dependence of the growth of structure as a test of General
  Relativity}},\ }\href@noop {} {\  (\bibinfo {year} {2019})},\ \Eprint
  {https://arxiv.org/abs/1906.02217} {arXiv:1906.02217 [astro-ph.CO]}
  \BibitemShut {NoStop}%
\bibitem [{\citenamefont {Arjona}\ and\ \citenamefont
  {Nesseris}(2020{\natexlab{a}})}]{Arjona:2019fwb}%
  \BibitemOpen
  \bibfield  {author} {\bibinfo {author} {\bibfnamefont {R.}~\bibnamefont
  {Arjona}}\ and\ \bibinfo {author} {\bibfnamefont {S.}~\bibnamefont
  {Nesseris}},\ }\bibfield  {title} {\bibinfo {title} {{What can Machine
  Learning tell us about the background expansion of the Universe?}},\ }\href
  {https://doi.org/10.1103/PhysRevD.101.123525} {\bibfield  {journal} {\bibinfo
   {journal} {Phys. Rev. D}\ }\textbf {\bibinfo {volume} {101}},\ \bibinfo
  {pages} {123525} (\bibinfo {year} {2020}{\natexlab{a}})},\ \Eprint
  {https://arxiv.org/abs/1910.01529} {arXiv:1910.01529 [astro-ph.CO]}
  \BibitemShut {NoStop}%
\bibitem [{\citenamefont {Seikel}\ \emph {et~al.}(2012)\citenamefont {Seikel},
  \citenamefont {Yahya}, \citenamefont {Maartens},\ and\ \citenamefont
  {Clarkson}}]{Seikel:2012cs}%
  \BibitemOpen
  \bibfield  {author} {\bibinfo {author} {\bibfnamefont {M.}~\bibnamefont
  {Seikel}}, \bibinfo {author} {\bibfnamefont {S.}~\bibnamefont {Yahya}},
  \bibinfo {author} {\bibfnamefont {R.}~\bibnamefont {Maartens}},\ and\
  \bibinfo {author} {\bibfnamefont {C.}~\bibnamefont {Clarkson}},\ }\bibfield
  {title} {\bibinfo {title} {{Using H(z) data as a probe of the concordance
  model}},\ }\href {https://doi.org/10.1103/PhysRevD.86.083001} {\bibfield
  {journal} {\bibinfo  {journal} {Phys. Rev. D}\ }\textbf {\bibinfo {volume}
  {86}},\ \bibinfo {pages} {083001} (\bibinfo {year} {2012})},\ \Eprint
  {https://arxiv.org/abs/1205.3431} {arXiv:1205.3431 [astro-ph.CO]}
  \BibitemShut {NoStop}%
\bibitem [{\citenamefont {Weinberg}(2008)}]{Weinberg:2008zzc}%
  \BibitemOpen
  \bibfield  {author} {\bibinfo {author} {\bibfnamefont {S.}~\bibnamefont
  {Weinberg}},\ }\href@noop {} {\emph {\bibinfo {title} {{Cosmology}}}}\
  (\bibinfo {year} {2008})\BibitemShut {NoStop}%
\bibitem [{\citenamefont {Maartens}(2011)}]{Maartens:2011yx}%
  \BibitemOpen
  \bibfield  {author} {\bibinfo {author} {\bibfnamefont {R.}~\bibnamefont
  {Maartens}},\ }\bibfield  {title} {\bibinfo {title} {{Is the Universe
  homogeneous?}},\ }\href {https://doi.org/10.1098/rsta.2011.0289} {\bibfield
  {journal} {\bibinfo  {journal} {Phil. Trans. Roy. Soc. Lond. A}\ }\textbf
  {\bibinfo {volume} {369}},\ \bibinfo {pages} {5115} (\bibinfo {year}
  {2011})},\ \Eprint {https://arxiv.org/abs/1104.1300} {arXiv:1104.1300
  [astro-ph.CO]} \BibitemShut {NoStop}%
\bibitem [{\citenamefont {Eisenstein}\ and\ \citenamefont
  {Hu}(1998)}]{Eisenstein:1997ik}%
  \BibitemOpen
  \bibfield  {author} {\bibinfo {author} {\bibfnamefont {D.~J.}\ \bibnamefont
  {Eisenstein}}\ and\ \bibinfo {author} {\bibfnamefont {W.}~\bibnamefont
  {Hu}},\ }\bibfield  {title} {\bibinfo {title} {{Baryonic features in the
  matter transfer function}},\ }\href {https://doi.org/10.1086/305424}
  {\bibfield  {journal} {\bibinfo  {journal} {Astrophys. J.}\ }\textbf
  {\bibinfo {volume} {496}},\ \bibinfo {pages} {605} (\bibinfo {year}
  {1998})},\ \Eprint {https://arxiv.org/abs/astro-ph/9709112}
  {arXiv:astro-ph/9709112 [astro-ph]} \BibitemShut {NoStop}%
\bibitem [{\citenamefont {Sanchez}\ \emph {et~al.}(2013)\citenamefont
  {Sanchez}, \citenamefont {Alonso}, \citenamefont {Sanchez}, \citenamefont
  {Garcia-Bellido},\ and\ \citenamefont {Sevilla}}]{Sanchez:2012eh}%
  \BibitemOpen
  \bibfield  {author} {\bibinfo {author} {\bibfnamefont {E.}~\bibnamefont
  {Sanchez}}, \bibinfo {author} {\bibfnamefont {D.}~\bibnamefont {Alonso}},
  \bibinfo {author} {\bibfnamefont {F.}~\bibnamefont {Sanchez}}, \bibinfo
  {author} {\bibfnamefont {J.}~\bibnamefont {Garcia-Bellido}},\ and\ \bibinfo
  {author} {\bibfnamefont {I.}~\bibnamefont {Sevilla}},\ }\bibfield  {title}
  {\bibinfo {title} {{Precise Measurement of the Radial Baryon Acoustic
  Oscillation Scales in Galaxy Redshift Surveys}},\ }\href
  {https://doi.org/10.1093/mnras/stt1146} {\bibfield  {journal} {\bibinfo
  {journal} {Mon. Not. Roy. Astron. Soc.}\ }\textbf {\bibinfo {volume} {434}},\
  \bibinfo {pages} {2008} (\bibinfo {year} {2013})},\ \Eprint
  {https://arxiv.org/abs/1210.6446} {arXiv:1210.6446 [astro-ph.CO]}
  \BibitemShut {NoStop}%
\bibitem [{\citenamefont {Gaztanaga}\ \emph {et~al.}(2009)\citenamefont
  {Gaztanaga}, \citenamefont {Cabre},\ and\ \citenamefont
  {Hui}}]{Gaztanaga:2008xz}%
  \BibitemOpen
  \bibfield  {author} {\bibinfo {author} {\bibfnamefont {E.}~\bibnamefont
  {Gaztanaga}}, \bibinfo {author} {\bibfnamefont {A.}~\bibnamefont {Cabre}},\
  and\ \bibinfo {author} {\bibfnamefont {L.}~\bibnamefont {Hui}},\ }\bibfield
  {title} {\bibinfo {title} {{Clustering of Luminous Red Galaxies IV: Baryon
  Acoustic Peak in the Line-of-Sight Direction and a Direct Measurement of
  H(z)}},\ }\href {https://doi.org/10.1111/j.1365-2966.2009.15405.x} {\bibfield
   {journal} {\bibinfo  {journal} {Mon. Not. Roy. Astron. Soc.}\ }\textbf
  {\bibinfo {volume} {399}},\ \bibinfo {pages} {1663} (\bibinfo {year}
  {2009})},\ \Eprint {https://arxiv.org/abs/0807.3551} {arXiv:0807.3551
  [astro-ph]} \BibitemShut {NoStop}%
\bibitem [{\citenamefont {Jimenez}\ and\ \citenamefont
  {Loeb}(2002)}]{Jimenez:2001gg}%
  \BibitemOpen
  \bibfield  {author} {\bibinfo {author} {\bibfnamefont {R.}~\bibnamefont
  {Jimenez}}\ and\ \bibinfo {author} {\bibfnamefont {A.}~\bibnamefont {Loeb}},\
  }\bibfield  {title} {\bibinfo {title} {{Constraining cosmological parameters
  based on relative galaxy ages}},\ }\href {https://doi.org/10.1086/340549}
  {\bibfield  {journal} {\bibinfo  {journal} {Astrophys. J.}\ }\textbf
  {\bibinfo {volume} {573}},\ \bibinfo {pages} {37} (\bibinfo {year} {2002})},\
  \Eprint {https://arxiv.org/abs/astro-ph/0106145} {arXiv:astro-ph/0106145
  [astro-ph]} \BibitemShut {NoStop}%
\bibitem [{\citenamefont {Moresco}\ \emph {et~al.}(2016)\citenamefont
  {Moresco}, \citenamefont {Pozzetti}, \citenamefont {Cimatti}, \citenamefont
  {Jimenez}, \citenamefont {Maraston}, \citenamefont {Verde}, \citenamefont
  {Thomas}, \citenamefont {Citro}, \citenamefont {Tojeiro},\ and\ \citenamefont
  {Wilkinson}}]{Moresco:2016mzx}%
  \BibitemOpen
  \bibfield  {author} {\bibinfo {author} {\bibfnamefont {M.}~\bibnamefont
  {Moresco}}, \bibinfo {author} {\bibfnamefont {L.}~\bibnamefont {Pozzetti}},
  \bibinfo {author} {\bibfnamefont {A.}~\bibnamefont {Cimatti}}, \bibinfo
  {author} {\bibfnamefont {R.}~\bibnamefont {Jimenez}}, \bibinfo {author}
  {\bibfnamefont {C.}~\bibnamefont {Maraston}}, \bibinfo {author}
  {\bibfnamefont {L.}~\bibnamefont {Verde}}, \bibinfo {author} {\bibfnamefont
  {D.}~\bibnamefont {Thomas}}, \bibinfo {author} {\bibfnamefont
  {A.}~\bibnamefont {Citro}}, \bibinfo {author} {\bibfnamefont
  {R.}~\bibnamefont {Tojeiro}},\ and\ \bibinfo {author} {\bibfnamefont
  {D.}~\bibnamefont {Wilkinson}},\ }\bibfield  {title} {\bibinfo {title} {{A
  6\% measurement of the Hubble parameter at $z\sim0.45$: direct evidence of
  the epoch of cosmic re-acceleration}},\ }\href
  {https://doi.org/10.1088/1475-7516/2016/05/014} {\bibfield  {journal}
  {\bibinfo  {journal} {JCAP}\ }\textbf {\bibinfo {volume} {1605}}\bibfield
  {number} {\bibinfo  {number} { (05)},\ \bibinfo {pages} {014}},\ }\Eprint
  {https://arxiv.org/abs/1601.01701} {arXiv:1601.01701 [astro-ph.CO]}
  \BibitemShut {NoStop}%
\bibitem [{\citenamefont {Zhang}\ \emph {et~al.}(2014)\citenamefont {Zhang},
  \citenamefont {Zhang}, \citenamefont {Yuan}, \citenamefont {Zhang},\ and\
  \citenamefont {Sun}}]{Zhang:2012mp}%
  \BibitemOpen
  \bibfield  {author} {\bibinfo {author} {\bibfnamefont {C.}~\bibnamefont
  {Zhang}}, \bibinfo {author} {\bibfnamefont {H.}~\bibnamefont {Zhang}},
  \bibinfo {author} {\bibfnamefont {S.}~\bibnamefont {Yuan}}, \bibinfo {author}
  {\bibfnamefont {T.-J.}\ \bibnamefont {Zhang}},\ and\ \bibinfo {author}
  {\bibfnamefont {Y.-C.}\ \bibnamefont {Sun}},\ }\bibfield  {title} {\bibinfo
  {title} {{Four new observational $H(z)$ data from luminous red galaxies in
  the Sloan Digital Sky Survey data release seven}},\ }\href
  {https://doi.org/10.1088/1674-4527/14/10/002} {\bibfield  {journal} {\bibinfo
   {journal} {Res. Astron. Astrophys.}\ }\textbf {\bibinfo {volume} {14}},\
  \bibinfo {pages} {1221} (\bibinfo {year} {2014})},\ \Eprint
  {https://arxiv.org/abs/1207.4541} {arXiv:1207.4541 [astro-ph.CO]}
  \BibitemShut {NoStop}%
\bibitem [{\citenamefont {Stern}\ \emph {et~al.}(2010)\citenamefont {Stern},
  \citenamefont {Jimenez}, \citenamefont {Verde}, \citenamefont
  {Kamionkowski},\ and\ \citenamefont {Stanford}}]{STERN:2009EP}%
  \BibitemOpen
  \bibfield  {author} {\bibinfo {author} {\bibfnamefont {D.}~\bibnamefont
  {Stern}}, \bibinfo {author} {\bibfnamefont {R.}~\bibnamefont {Jimenez}},
  \bibinfo {author} {\bibfnamefont {L.}~\bibnamefont {Verde}}, \bibinfo
  {author} {\bibfnamefont {M.}~\bibnamefont {Kamionkowski}},\ and\ \bibinfo
  {author} {\bibfnamefont {S.~A.}\ \bibnamefont {Stanford}},\ }\bibfield
  {title} {\bibinfo {title} {{Cosmic Chronometers: Constraining the Equation of
  State of Dark Energy. I: H(z) Measurements}},\ }\href
  {https://doi.org/10.1088/1475-7516/2010/02/008} {\bibfield  {journal}
  {\bibinfo  {journal} {JCAP}\ }\textbf {\bibinfo {volume} {1002}},\ \bibinfo
  {pages} {008}},\ \Eprint {https://arxiv.org/abs/0907.3149} {arXiv:0907.3149
  [astro-ph.CO]} \BibitemShut {NoStop}%
\bibitem [{\citenamefont {Moresco}\ \emph {et~al.}(2012)\citenamefont {Moresco}
  \emph {et~al.}}]{MORESCO:2012JH}%
  \BibitemOpen
  \bibfield  {author} {\bibinfo {author} {\bibfnamefont {M.}~\bibnamefont
  {Moresco}} \emph {et~al.},\ }\bibfield  {title} {\bibinfo {title} {{Improved
  constraints on the expansion rate of the Universe up to z~1.1 from the
  spectroscopic evolution of cosmic chronometers}},\ }\href
  {https://doi.org/10.1088/1475-7516/2012/08/006} {\bibfield  {journal}
  {\bibinfo  {journal} {JCAP}\ }\textbf {\bibinfo {volume} {1208}},\ \bibinfo
  {pages} {006}},\ \Eprint {https://arxiv.org/abs/1201.3609} {arXiv:1201.3609
  [astro-ph.CO]} \BibitemShut {NoStop}%
\bibitem [{\citenamefont {Chuang}\ and\ \citenamefont
  {Wang}(2013)}]{Chuang:2012qt}%
  \BibitemOpen
  \bibfield  {author} {\bibinfo {author} {\bibfnamefont {C.-H.}\ \bibnamefont
  {Chuang}}\ and\ \bibinfo {author} {\bibfnamefont {Y.}~\bibnamefont {Wang}},\
  }\bibfield  {title} {\bibinfo {title} {{Modeling the Anisotropic Two-Point
  Galaxy Correlation Function on Small Scales and Improved Measurements of
  $H(z)$, $D_A(z)$, and $\beta(z)$ from the Sloan Digital Sky Survey DR7
  Luminous Red Galaxies}},\ }\href {https://doi.org/10.1093/mnras/stt1290}
  {\bibfield  {journal} {\bibinfo  {journal} {Mon. Not. Roy. Astron. Soc.}\
  }\textbf {\bibinfo {volume} {435}},\ \bibinfo {pages} {255} (\bibinfo {year}
  {2013})},\ \Eprint {https://arxiv.org/abs/1209.0210} {arXiv:1209.0210
  [astro-ph.CO]} \BibitemShut {NoStop}%
\bibitem [{\citenamefont {Blake}\ \emph {et~al.}(2012)\citenamefont {Blake}
  \emph {et~al.}}]{Blake:2012pj}%
  \BibitemOpen
  \bibfield  {author} {\bibinfo {author} {\bibfnamefont {C.}~\bibnamefont
  {Blake}} \emph {et~al.},\ }\bibfield  {title} {\bibinfo {title} {{The WiggleZ
  Dark Energy Survey: Joint measurements of the expansion and growth history at
  z < 1}},\ }\href {https://doi.org/10.1111/j.1365-2966.2012.21473.x}
  {\bibfield  {journal} {\bibinfo  {journal} {Mon. Not. Roy. Astron. Soc.}\
  }\textbf {\bibinfo {volume} {425}},\ \bibinfo {pages} {405} (\bibinfo {year}
  {2012})},\ \Eprint {https://arxiv.org/abs/1204.3674} {arXiv:1204.3674
  [astro-ph.CO]} \BibitemShut {NoStop}%
\bibitem [{\citenamefont {Anderson}\ \emph {et~al.}(2014)\citenamefont
  {Anderson} \emph {et~al.}}]{Anderson:2013zyy}%
  \BibitemOpen
  \bibfield  {author} {\bibinfo {author} {\bibfnamefont {L.}~\bibnamefont
  {Anderson}} \emph {et~al.} (\bibinfo {collaboration} {BOSS}),\ }\bibfield
  {title} {\bibinfo {title} {{The clustering of galaxies in the SDSS-III Baryon
  Oscillation Spectroscopic Survey: baryon acoustic oscillations in the Data
  Releases 10 and 11 Galaxy samples}},\ }\href
  {https://doi.org/10.1093/mnras/stu523} {\bibfield  {journal} {\bibinfo
  {journal} {Mon. Not. Roy. Astron. Soc.}\ }\textbf {\bibinfo {volume} {441}},\
  \bibinfo {pages} {24} (\bibinfo {year} {2014})},\ \Eprint
  {https://arxiv.org/abs/1312.4877} {arXiv:1312.4877 [astro-ph.CO]}
  \BibitemShut {NoStop}%
\bibitem [{\citenamefont {Delubac}\ \emph {et~al.}(2015)\citenamefont {Delubac}
  \emph {et~al.}}]{Delubac:2014aqe}%
  \BibitemOpen
  \bibfield  {author} {\bibinfo {author} {\bibfnamefont {T.}~\bibnamefont
  {Delubac}} \emph {et~al.} (\bibinfo {collaboration} {BOSS}),\ }\bibfield
  {title} {\bibinfo {title} {{Baryon acoustic oscillations in the Lya forest of
  BOSS DR11 quasars}},\ }\href {https://doi.org/10.1051/0004-6361/201423969}
  {\bibfield  {journal} {\bibinfo  {journal} {Astron. Astrophys.}\ }\textbf
  {\bibinfo {volume} {574}},\ \bibinfo {pages} {A59} (\bibinfo {year}
  {2015})},\ \Eprint {https://arxiv.org/abs/1404.1801} {arXiv:1404.1801
  [astro-ph.CO]} \BibitemShut {NoStop}%
\bibitem [{\citenamefont {Guo}\ and\ \citenamefont
  {Zhang}(2016)}]{Guo:2015gpa}%
  \BibitemOpen
  \bibfield  {author} {\bibinfo {author} {\bibfnamefont {R.-Y.}\ \bibnamefont
  {Guo}}\ and\ \bibinfo {author} {\bibfnamefont {X.}~\bibnamefont {Zhang}},\
  }\bibfield  {title} {\bibinfo {title} {{Constraining dark energy with Hubble
  parameter measurements: an analysis including future redshift-drift
  observations}},\ }\href {https://doi.org/10.1140/epjc/s10052-016-4016-x}
  {\bibfield  {journal} {\bibinfo  {journal} {Eur. Phys. J.}\ }\textbf
  {\bibinfo {volume} {C76}},\ \bibinfo {pages} {163} (\bibinfo {year}
  {2016})},\ \Eprint {https://arxiv.org/abs/1512.07703} {arXiv:1512.07703
  [astro-ph.CO]} \BibitemShut {NoStop}%
\bibitem [{\citenamefont {Yu}\ \emph {et~al.}(2018)\citenamefont {Yu},
  \citenamefont {Ratra},\ and\ \citenamefont {Wang}}]{Yu:2017iju}%
  \BibitemOpen
  \bibfield  {author} {\bibinfo {author} {\bibfnamefont {H.}~\bibnamefont
  {Yu}}, \bibinfo {author} {\bibfnamefont {B.}~\bibnamefont {Ratra}},\ and\
  \bibinfo {author} {\bibfnamefont {F.-Y.}\ \bibnamefont {Wang}},\ }\bibfield
  {title} {\bibinfo {title} {{Hubble Parameter and Baryon Acoustic Oscillation
  Measurement Constraints on the Hubble Constant, the Deviation from the
  Spatially Flat LCDM Model, the Deceleration–Acceleration Transition
  Redshift, and Spatial Curvature}},\ }\href
  {https://doi.org/10.3847/1538-4357/aab0a2} {\bibfield  {journal} {\bibinfo
  {journal} {Astrophys. J.}\ }\textbf {\bibinfo {volume} {856}},\ \bibinfo
  {pages} {3} (\bibinfo {year} {2018})},\ \Eprint
  {https://arxiv.org/abs/1711.03437} {arXiv:1711.03437 [astro-ph.CO]}
  \BibitemShut {NoStop}%
\bibitem [{\citenamefont {Beutler}\ \emph {et~al.}(2011)\citenamefont
  {Beutler}, \citenamefont {Blake}, \citenamefont {Colless}, \citenamefont
  {Jones}, \citenamefont {Staveley-Smith}, \citenamefont {Campbell},
  \citenamefont {Parker}, \citenamefont {Saunders},\ and\ \citenamefont
  {Watson}}]{Beutler:2011hx}%
  \BibitemOpen
  \bibfield  {author} {\bibinfo {author} {\bibfnamefont {F.}~\bibnamefont
  {Beutler}}, \bibinfo {author} {\bibfnamefont {C.}~\bibnamefont {Blake}},
  \bibinfo {author} {\bibfnamefont {M.}~\bibnamefont {Colless}}, \bibinfo
  {author} {\bibfnamefont {D.~H.}\ \bibnamefont {Jones}}, \bibinfo {author}
  {\bibfnamefont {L.}~\bibnamefont {Staveley-Smith}}, \bibinfo {author}
  {\bibfnamefont {L.}~\bibnamefont {Campbell}}, \bibinfo {author}
  {\bibfnamefont {Q.}~\bibnamefont {Parker}}, \bibinfo {author} {\bibfnamefont
  {W.}~\bibnamefont {Saunders}},\ and\ \bibinfo {author} {\bibfnamefont
  {F.}~\bibnamefont {Watson}},\ }\bibfield  {title} {\bibinfo {title} {{The 6dF
  Galaxy Survey: Baryon Acoustic Oscillations and the Local Hubble Constant}},\
  }\href {https://doi.org/10.1111/j.1365-2966.2011.19250.x} {\bibfield
  {journal} {\bibinfo  {journal} {Mon. Not. Roy. Astron. Soc.}\ }\textbf
  {\bibinfo {volume} {416}},\ \bibinfo {pages} {3017} (\bibinfo {year}
  {2011})},\ \Eprint {https://arxiv.org/abs/1106.3366} {arXiv:1106.3366
  [astro-ph.CO]} \BibitemShut {NoStop}%
\bibitem [{\citenamefont {Xu}\ \emph {et~al.}(2012)\citenamefont {Xu},
  \citenamefont {Padmanabhan}, \citenamefont {Eisenstein}, \citenamefont
  {Mehta},\ and\ \citenamefont {Cuesta}}]{Xu:2012hg}%
  \BibitemOpen
  \bibfield  {author} {\bibinfo {author} {\bibfnamefont {X.}~\bibnamefont
  {Xu}}, \bibinfo {author} {\bibfnamefont {N.}~\bibnamefont {Padmanabhan}},
  \bibinfo {author} {\bibfnamefont {D.~J.}\ \bibnamefont {Eisenstein}},
  \bibinfo {author} {\bibfnamefont {K.~T.}\ \bibnamefont {Mehta}},\ and\
  \bibinfo {author} {\bibfnamefont {A.~J.}\ \bibnamefont {Cuesta}},\ }\bibfield
   {title} {\bibinfo {title} {{A 2\% Distance to z=0.35 by Reconstructing
  Baryon Acoustic Oscillations - II: Fitting Techniques}},\ }\href
  {https://doi.org/10.1111/j.1365-2966.2012.21573.x} {\bibfield  {journal}
  {\bibinfo  {journal} {Mon. Not. Roy. Astron. Soc.}\ }\textbf {\bibinfo
  {volume} {427}},\ \bibinfo {pages} {2146} (\bibinfo {year} {2012})},\ \Eprint
  {https://arxiv.org/abs/1202.0091} {arXiv:1202.0091 [astro-ph.CO]}
  \BibitemShut {NoStop}%
\bibitem [{\citenamefont {Ross}\ \emph {et~al.}(2015)\citenamefont {Ross},
  \citenamefont {Samushia}, \citenamefont {Howlett}, \citenamefont {Percival},
  \citenamefont {Burden},\ and\ \citenamefont {Manera}}]{Ross:2014qpa}%
  \BibitemOpen
  \bibfield  {author} {\bibinfo {author} {\bibfnamefont {A.~J.}\ \bibnamefont
  {Ross}}, \bibinfo {author} {\bibfnamefont {L.}~\bibnamefont {Samushia}},
  \bibinfo {author} {\bibfnamefont {C.}~\bibnamefont {Howlett}}, \bibinfo
  {author} {\bibfnamefont {W.~J.}\ \bibnamefont {Percival}}, \bibinfo {author}
  {\bibfnamefont {A.}~\bibnamefont {Burden}},\ and\ \bibinfo {author}
  {\bibfnamefont {M.}~\bibnamefont {Manera}},\ }\bibfield  {title} {\bibinfo
  {title} {{The clustering of the SDSS DR7 main Galaxy sample I. A 4 per cent
  distance measure at $z = 0.15$}},\ }\href
  {https://doi.org/10.1093/mnras/stv154} {\bibfield  {journal} {\bibinfo
  {journal} {Mon. Not. Roy. Astron. Soc.}\ }\textbf {\bibinfo {volume} {449}},\
  \bibinfo {pages} {835} (\bibinfo {year} {2015})},\ \Eprint
  {https://arxiv.org/abs/1409.3242} {arXiv:1409.3242 [astro-ph.CO]}
  \BibitemShut {NoStop}%
\bibitem [{\citenamefont {Gil-Marin}\ \emph {et~al.}(2016)\citenamefont
  {Gil-Marin} \emph {et~al.}}]{Gil-Marin:2015nqa}%
  \BibitemOpen
  \bibfield  {author} {\bibinfo {author} {\bibfnamefont {H.}~\bibnamefont
  {Gil-Marin}} \emph {et~al.},\ }\bibfield  {title} {\bibinfo {title} {{The
  clustering of galaxies in the SDSS-III Baryon Oscillation Spectroscopic
  Survey: BAO measurement from the LOS-dependent power spectrum of DR12 BOSS
  galaxies}},\ }\href {https://doi.org/10.1093/mnras/stw1264} {\bibfield
  {journal} {\bibinfo  {journal} {Mon. Not. Roy. Astron. Soc.}\ }\textbf
  {\bibinfo {volume} {460}},\ \bibinfo {pages} {4210} (\bibinfo {year}
  {2016})},\ \Eprint {https://arxiv.org/abs/1509.06373} {arXiv:1509.06373
  [astro-ph.CO]} \BibitemShut {NoStop}%
\bibitem [{\citenamefont {Abbott}\ \emph {et~al.}(2019)\citenamefont {Abbott}
  \emph {et~al.}}]{Abbott:2017wcz}%
  \BibitemOpen
  \bibfield  {author} {\bibinfo {author} {\bibfnamefont {T.~M.~C.}\
  \bibnamefont {Abbott}} \emph {et~al.} (\bibinfo {collaboration} {DES}),\
  }\bibfield  {title} {\bibinfo {title} {{Dark Energy Survey Year 1 Results:
  Measurement of the Baryon Acoustic Oscillation scale in the distribution of
  galaxies to redshift 1}},\ }\href {https://doi.org/10.1093/mnras/sty3351}
  {\bibfield  {journal} {\bibinfo  {journal} {Mon. Not. Roy. Astron. Soc.}\
  }\textbf {\bibinfo {volume} {483}},\ \bibinfo {pages} {4866} (\bibinfo {year}
  {2019})},\ \Eprint {https://arxiv.org/abs/1712.06209} {arXiv:1712.06209
  [astro-ph.CO]} \BibitemShut {NoStop}%
\bibitem [{\citenamefont {Blomqvist}\ \emph {et~al.}(2019)\citenamefont
  {Blomqvist} \emph {et~al.}}]{Blomqvist:2019rah}%
  \BibitemOpen
  \bibfield  {author} {\bibinfo {author} {\bibfnamefont {M.}~\bibnamefont
  {Blomqvist}} \emph {et~al.},\ }\bibfield  {title} {\bibinfo {title} {{Baryon
  acoustic oscillations from the cross-correlation of Ly$\alpha$ absorption and
  quasars in eBOSS DR14}},\ }\href
  {https://doi.org/10.1051/0004-6361/201935641} {\bibfield  {journal} {\bibinfo
   {journal} {Astron. Astrophys.}\ }\textbf {\bibinfo {volume} {629}},\
  \bibinfo {pages} {A86} (\bibinfo {year} {2019})},\ \Eprint
  {https://arxiv.org/abs/1904.03430} {arXiv:1904.03430 [astro-ph.CO]}
  \BibitemShut {NoStop}%
\bibitem [{\citenamefont {Bautista}\ \emph {et~al.}(2018)\citenamefont
  {Bautista} \emph {et~al.}}]{Bautista:2017wwp}%
  \BibitemOpen
  \bibfield  {author} {\bibinfo {author} {\bibfnamefont {J.~E.}\ \bibnamefont
  {Bautista}} \emph {et~al.},\ }\bibfield  {title} {\bibinfo {title} {{The
  SDSS-IV extended Baryon Oscillation Spectroscopic Survey: Baryon Acoustic
  Oscillations at redshift of 0.72 with the DR14 Luminous Red Galaxy Sample}},\
  }\href {https://doi.org/10.3847/1538-4357/aacea5} {\bibfield  {journal}
  {\bibinfo  {journal} {Astrophys. J.}\ }\textbf {\bibinfo {volume} {863}},\
  \bibinfo {pages} {110} (\bibinfo {year} {2018})},\ \Eprint
  {https://arxiv.org/abs/1712.08064} {arXiv:1712.08064 [astro-ph.CO]}
  \BibitemShut {NoStop}%
\bibitem [{\citenamefont {Ata}\ \emph {et~al.}(2018)\citenamefont {Ata} \emph
  {et~al.}}]{Ata:2017dya}%
  \BibitemOpen
  \bibfield  {author} {\bibinfo {author} {\bibfnamefont {M.}~\bibnamefont
  {Ata}} \emph {et~al.},\ }\bibfield  {title} {\bibinfo {title} {{The
  clustering of the SDSS-IV extended Baryon Oscillation Spectroscopic Survey
  DR14 quasar sample: first measurement of baryon acoustic oscillations between
  redshift 0.8 and 2.2}},\ }\href {https://doi.org/10.1093/mnras/stx2630}
  {\bibfield  {journal} {\bibinfo  {journal} {Mon. Not. Roy. Astron. Soc.}\
  }\textbf {\bibinfo {volume} {473}},\ \bibinfo {pages} {4773} (\bibinfo {year}
  {2018})},\ \Eprint {https://arxiv.org/abs/1705.06373} {arXiv:1705.06373
  [astro-ph.CO]} \BibitemShut {NoStop}%
\bibitem [{\citenamefont {Alam}\ \emph {et~al.}(2016)\citenamefont {Alam},
  \citenamefont {Ho},\ and\ \citenamefont {Silvestri}}]{Alam:2015rsa}%
  \BibitemOpen
  \bibfield  {author} {\bibinfo {author} {\bibfnamefont {S.}~\bibnamefont
  {Alam}}, \bibinfo {author} {\bibfnamefont {S.}~\bibnamefont {Ho}},\ and\
  \bibinfo {author} {\bibfnamefont {A.}~\bibnamefont {Silvestri}},\ }\bibfield
  {title} {\bibinfo {title} {{Testing deviations from \ensuremath{\Lambda}CDM
  with growth rate measurements from six large-scale structure surveys at $z =
  $0.06\textendash{}1}},\ }\href {https://doi.org/10.1093/mnras/stv2935}
  {\bibfield  {journal} {\bibinfo  {journal} {Mon. Not. Roy. Astron. Soc.}\
  }\textbf {\bibinfo {volume} {456}},\ \bibinfo {pages} {3743} (\bibinfo {year}
  {2016})},\ \Eprint {https://arxiv.org/abs/1509.05034} {arXiv:1509.05034
  [astro-ph.CO]} \BibitemShut {NoStop}%
\bibitem [{\citenamefont {Alam}\ \emph {et~al.}(2020)\citenamefont {Alam} \emph
  {et~al.}}]{Alam:2020sor}%
  \BibitemOpen
  \bibfield  {author} {\bibinfo {author} {\bibfnamefont {S.}~\bibnamefont
  {Alam}} \emph {et~al.} (\bibinfo {collaboration} {eBOSS}),\ }\bibfield
  {title} {\bibinfo {title} {{The Completed SDSS-IV extended Baryon Oscillation
  Spectroscopic Survey: Cosmological Implications from two Decades of
  Spectroscopic Surveys at the Apache Point observatory}},\ }\href@noop {} {\
  (\bibinfo {year} {2020})},\ \Eprint {https://arxiv.org/abs/2007.08991}
  {arXiv:2007.08991 [astro-ph.CO]} \BibitemShut {NoStop}%
\bibitem [{\citenamefont {Nunes}\ \emph {et~al.}(2020)\citenamefont {Nunes},
  \citenamefont {Yadav}, \citenamefont {Jesus},\ and\ \citenamefont
  {Bernui}}]{Nunes:2020hzy}%
  \BibitemOpen
  \bibfield  {author} {\bibinfo {author} {\bibfnamefont {R.~C.}\ \bibnamefont
  {Nunes}}, \bibinfo {author} {\bibfnamefont {S.~K.}\ \bibnamefont {Yadav}},
  \bibinfo {author} {\bibfnamefont {J.~F.}\ \bibnamefont {Jesus}},\ and\
  \bibinfo {author} {\bibfnamefont {A.}~\bibnamefont {Bernui}},\ }\bibfield
  {title} {\bibinfo {title} {{Cosmological parameter analyses using transversal
  BAO data}},\ }\href {https://doi.org/10.1093/mnras/staa2036} {\bibfield
  {journal} {\bibinfo  {journal} {Mon. Not. Roy. Astron. Soc.}\ }\textbf
  {\bibinfo {volume} {497}},\ \bibinfo {pages} {2133} (\bibinfo {year}
  {2020})},\ \Eprint {https://arxiv.org/abs/2002.09293} {arXiv:2002.09293
  [astro-ph.CO]} \BibitemShut {NoStop}%
\bibitem [{\citenamefont {Sanchez}\ \emph {et~al.}(2011)\citenamefont
  {Sanchez}, \citenamefont {Carnero}, \citenamefont {Garcia-Bellido},
  \citenamefont {Gaztanaga}, \citenamefont {de~Simoni}, \citenamefont {Crocce},
  \citenamefont {Cabre}, \citenamefont {Fosalba},\ and\ \citenamefont
  {Alonso}}]{Sanchez:2010zg}%
  \BibitemOpen
  \bibfield  {author} {\bibinfo {author} {\bibfnamefont {E.}~\bibnamefont
  {Sanchez}}, \bibinfo {author} {\bibfnamefont {A.}~\bibnamefont {Carnero}},
  \bibinfo {author} {\bibfnamefont {J.}~\bibnamefont {Garcia-Bellido}},
  \bibinfo {author} {\bibfnamefont {E.}~\bibnamefont {Gaztanaga}}, \bibinfo
  {author} {\bibfnamefont {F.}~\bibnamefont {de~Simoni}}, \bibinfo {author}
  {\bibfnamefont {M.}~\bibnamefont {Crocce}}, \bibinfo {author} {\bibfnamefont
  {A.}~\bibnamefont {Cabre}}, \bibinfo {author} {\bibfnamefont
  {P.}~\bibnamefont {Fosalba}},\ and\ \bibinfo {author} {\bibfnamefont
  {D.}~\bibnamefont {Alonso}},\ }\bibfield  {title} {\bibinfo {title} {{Tracing
  The Sound Horizon Scale With Photometric Redshift Surveys}},\ }\href
  {https://doi.org/10.1111/j.1365-2966.2010.17679.x} {\bibfield  {journal}
  {\bibinfo  {journal} {Mon. Not. Roy. Astron. Soc.}\ }\textbf {\bibinfo
  {volume} {411}},\ \bibinfo {pages} {277} (\bibinfo {year} {2011})},\ \Eprint
  {https://arxiv.org/abs/1006.3226} {arXiv:1006.3226 [astro-ph.CO]}
  \BibitemShut {NoStop}%
\bibitem [{\citenamefont {de~Carvalho}\ \emph {et~al.}(2018)\citenamefont
  {de~Carvalho}, \citenamefont {Bernui}, \citenamefont {Carvalho},
  \citenamefont {Novaes},\ and\ \citenamefont {Xavier}}]{deCarvalho:2017xye}%
  \BibitemOpen
  \bibfield  {author} {\bibinfo {author} {\bibfnamefont {E.}~\bibnamefont
  {de~Carvalho}}, \bibinfo {author} {\bibfnamefont {A.}~\bibnamefont {Bernui}},
  \bibinfo {author} {\bibfnamefont {G.}~\bibnamefont {Carvalho}}, \bibinfo
  {author} {\bibfnamefont {C.}~\bibnamefont {Novaes}},\ and\ \bibinfo {author}
  {\bibfnamefont {H.}~\bibnamefont {Xavier}},\ }\bibfield  {title} {\bibinfo
  {title} {{Angular Baryon Acoustic Oscillation measure at $z=2.225$ from the
  SDSS quasar survey}},\ }\href {https://doi.org/10.1088/1475-7516/2018/04/064}
  {\bibfield  {journal} {\bibinfo  {journal} {JCAP}\ }\textbf {\bibinfo
  {volume} {04}},\ \bibinfo {pages} {064}},\ \Eprint
  {https://arxiv.org/abs/1709.00113} {arXiv:1709.00113 [astro-ph.CO]}
  \BibitemShut {NoStop}%
\bibitem [{\citenamefont {Carvalho}\ \emph {et~al.}(2020)\citenamefont
  {Carvalho}, \citenamefont {Bernui}, \citenamefont {Benetti}, \citenamefont
  {Carvalho}, \citenamefont {de~Carvalho},\ and\ \citenamefont
  {Alcaniz}}]{Carvalho:2017tuu}%
  \BibitemOpen
  \bibfield  {author} {\bibinfo {author} {\bibfnamefont {G.}~\bibnamefont
  {Carvalho}}, \bibinfo {author} {\bibfnamefont {A.}~\bibnamefont {Bernui}},
  \bibinfo {author} {\bibfnamefont {M.}~\bibnamefont {Benetti}}, \bibinfo
  {author} {\bibfnamefont {J.}~\bibnamefont {Carvalho}}, \bibinfo {author}
  {\bibfnamefont {E.}~\bibnamefont {de~Carvalho}},\ and\ \bibinfo {author}
  {\bibfnamefont {J.}~\bibnamefont {Alcaniz}},\ }\bibfield  {title} {\bibinfo
  {title} {{The transverse baryonic acoustic scale from the SDSS DR11
  galaxies}},\ }\href {https://doi.org/10.1016/j.astropartphys.2020.102432}
  {\bibfield  {journal} {\bibinfo  {journal} {Astropart. Phys.}\ }\textbf
  {\bibinfo {volume} {119}},\ \bibinfo {pages} {102432} (\bibinfo {year}
  {2020})},\ \Eprint {https://arxiv.org/abs/1709.00271} {arXiv:1709.00271
  [astro-ph.CO]} \BibitemShut {NoStop}%
\bibitem [{\citenamefont {Carvalho}\ \emph {et~al.}(2016)\citenamefont
  {Carvalho}, \citenamefont {Bernui}, \citenamefont {Benetti}, \citenamefont
  {Carvalho},\ and\ \citenamefont {Alcaniz}}]{Carvalho:2015ica}%
  \BibitemOpen
  \bibfield  {author} {\bibinfo {author} {\bibfnamefont {G.}~\bibnamefont
  {Carvalho}}, \bibinfo {author} {\bibfnamefont {A.}~\bibnamefont {Bernui}},
  \bibinfo {author} {\bibfnamefont {M.}~\bibnamefont {Benetti}}, \bibinfo
  {author} {\bibfnamefont {J.}~\bibnamefont {Carvalho}},\ and\ \bibinfo
  {author} {\bibfnamefont {J.}~\bibnamefont {Alcaniz}},\ }\bibfield  {title}
  {\bibinfo {title} {{Baryon Acoustic Oscillations from the SDSS DR10 galaxies
  angular correlation function}},\ }\href
  {https://doi.org/10.1103/PhysRevD.93.023530} {\bibfield  {journal} {\bibinfo
  {journal} {Phys. Rev. D}\ }\textbf {\bibinfo {volume} {93}},\ \bibinfo
  {pages} {023530} (\bibinfo {year} {2016})},\ \Eprint
  {https://arxiv.org/abs/1507.08972} {arXiv:1507.08972 [astro-ph.CO]}
  \BibitemShut {NoStop}%
\bibitem [{\citenamefont {Alcaniz}\ \emph {et~al.}(2017)\citenamefont
  {Alcaniz}, \citenamefont {Carvalho}, \citenamefont {Bernui}, \citenamefont
  {Carvalho},\ and\ \citenamefont {Benetti}}]{Alcaniz:2016ryy}%
  \BibitemOpen
  \bibfield  {author} {\bibinfo {author} {\bibfnamefont {J.~S.}\ \bibnamefont
  {Alcaniz}}, \bibinfo {author} {\bibfnamefont {G.~C.}\ \bibnamefont
  {Carvalho}}, \bibinfo {author} {\bibfnamefont {A.}~\bibnamefont {Bernui}},
  \bibinfo {author} {\bibfnamefont {J.~C.}\ \bibnamefont {Carvalho}},\ and\
  \bibinfo {author} {\bibfnamefont {M.}~\bibnamefont {Benetti}},\ }\bibinfo
  {title} {{Measuring baryon acoustic oscillations with angular two-point
  correlation function}}\ (\bibinfo {year} {2017})\ pp.\ \bibinfo {pages}
  {11--19},\ \Eprint {https://arxiv.org/abs/1611.08458} {arXiv:1611.08458
  [astro-ph.CO]} \BibitemShut {NoStop}%
\bibitem [{\citenamefont {Bogdanos}\ and\ \citenamefont
  {Nesseris}(2009)}]{Bogdanos:2009ib}%
  \BibitemOpen
  \bibfield  {author} {\bibinfo {author} {\bibfnamefont {C.}~\bibnamefont
  {Bogdanos}}\ and\ \bibinfo {author} {\bibfnamefont {S.}~\bibnamefont
  {Nesseris}},\ }\bibfield  {title} {\bibinfo {title} {{Genetic Algorithms and
  Supernovae Type Ia Analysis}},\ }\href
  {https://doi.org/10.1088/1475-7516/2009/05/006} {\bibfield  {journal}
  {\bibinfo  {journal} {JCAP}\ }\textbf {\bibinfo {volume} {0905}},\ \bibinfo
  {pages} {006}},\ \Eprint {https://arxiv.org/abs/0903.2805} {arXiv:0903.2805
  [astro-ph.CO]} \BibitemShut {NoStop}%
\bibitem [{\citenamefont {Nesseris}\ and\ \citenamefont
  {Garcia-Bellido}(2012)}]{Nesseris:2012tt}%
  \BibitemOpen
  \bibfield  {author} {\bibinfo {author} {\bibfnamefont {S.}~\bibnamefont
  {Nesseris}}\ and\ \bibinfo {author} {\bibfnamefont {J.}~\bibnamefont
  {Garcia-Bellido}},\ }\bibfield  {title} {\bibinfo {title} {{A new perspective
  on Dark Energy modeling via Genetic Algorithms}},\ }\href
  {https://doi.org/10.1088/1475-7516/2012/11/033} {\bibfield  {journal}
  {\bibinfo  {journal} {JCAP}\ }\textbf {\bibinfo {volume} {11}},\ \bibinfo
  {pages} {033}},\ \Eprint {https://arxiv.org/abs/1205.0364} {arXiv:1205.0364
  [astro-ph.CO]} \BibitemShut {NoStop}%
\bibitem [{\citenamefont {Nesseris}\ and\ \citenamefont
  {Garc\'\i{}a-Bellido}(2013)}]{Nesseris:2013bia}%
  \BibitemOpen
  \bibfield  {author} {\bibinfo {author} {\bibfnamefont {S.}~\bibnamefont
  {Nesseris}}\ and\ \bibinfo {author} {\bibfnamefont {J.}~\bibnamefont
  {Garc\'\i{}a-Bellido}},\ }\bibfield  {title} {\bibinfo {title} {{Comparative
  analysis of model-independent methods for exploring the nature of dark
  energy}},\ }\href {https://doi.org/10.1103/PhysRevD.88.063521} {\bibfield
  {journal} {\bibinfo  {journal} {Phys. Rev. D}\ }\textbf {\bibinfo {volume}
  {88}},\ \bibinfo {pages} {063521} (\bibinfo {year} {2013})},\ \Eprint
  {https://arxiv.org/abs/1306.4885} {arXiv:1306.4885 [astro-ph.CO]}
  \BibitemShut {NoStop}%
\bibitem [{\citenamefont {Arjona}\ and\ \citenamefont
  {Nesseris}(2020{\natexlab{b}})}]{Arjona:2020skf}%
  \BibitemOpen
  \bibfield  {author} {\bibinfo {author} {\bibfnamefont {R.}~\bibnamefont
  {Arjona}}\ and\ \bibinfo {author} {\bibfnamefont {S.}~\bibnamefont
  {Nesseris}},\ }\bibfield  {title} {\bibinfo {title} {{Machine Learning and
  cosmographic reconstructions of quintessence and the Swampland
  conjectures}},\ }\href@noop {} {\  (\bibinfo {year} {2020}{\natexlab{b}})},\
  \Eprint {https://arxiv.org/abs/2012.12202} {arXiv:2012.12202 [astro-ph.CO]}
  \BibitemShut {NoStop}%
\bibitem [{\citenamefont {Arjona}\ \emph {et~al.}(2021)\citenamefont {Arjona},
  \citenamefont {Lin}, \citenamefont {Nesseris},\ and\ \citenamefont
  {Tang}}]{Arjona:2020axn}%
  \BibitemOpen
  \bibfield  {author} {\bibinfo {author} {\bibfnamefont {R.}~\bibnamefont
  {Arjona}}, \bibinfo {author} {\bibfnamefont {H.-N.}\ \bibnamefont {Lin}},
  \bibinfo {author} {\bibfnamefont {S.}~\bibnamefont {Nesseris}},\ and\
  \bibinfo {author} {\bibfnamefont {L.}~\bibnamefont {Tang}},\ }\bibfield
  {title} {\bibinfo {title} {{Machine learning forecasts of the cosmic distance
  duality relation with strongly lensed gravitational wave events}},\ }\href
  {https://doi.org/10.1103/PhysRevD.103.103513} {\bibfield  {journal} {\bibinfo
   {journal} {Phys. Rev. D}\ }\textbf {\bibinfo {volume} {103}},\ \bibinfo
  {pages} {103513} (\bibinfo {year} {2021})},\ \Eprint
  {https://arxiv.org/abs/2011.02718} {arXiv:2011.02718 [astro-ph.CO]}
  \BibitemShut {NoStop}%
\bibitem [{\citenamefont {Arjona}(2020{\natexlab{b}})}]{Arjona:2020doi}%
  \BibitemOpen
  \bibfield  {author} {\bibinfo {author} {\bibfnamefont {R.}~\bibnamefont
  {Arjona}},\ }\bibfield  {title} {\bibinfo {title} {{Machine Learning meets
  the redshift evolution of the CMB Temperature}},\ }\href
  {https://doi.org/10.1088/1475-7516/2020/08/009} {\bibfield  {journal}
  {\bibinfo  {journal} {JCAP}\ }\textbf {\bibinfo {volume} {08}},\ \bibinfo
  {pages} {009}},\ \Eprint {https://arxiv.org/abs/2002.12700} {arXiv:2002.12700
  [astro-ph.CO]} \BibitemShut {NoStop}%
\bibitem [{\citenamefont {Arjona}\ and\ \citenamefont
  {Nesseris}(2020{\natexlab{c}})}]{Arjona:2020kco}%
  \BibitemOpen
  \bibfield  {author} {\bibinfo {author} {\bibfnamefont {R.}~\bibnamefont
  {Arjona}}\ and\ \bibinfo {author} {\bibfnamefont {S.}~\bibnamefont
  {Nesseris}},\ }\bibfield  {title} {\bibinfo {title} {{Hints of dark energy
  anisotropic stress using Machine Learning}},\ }\href@noop {} {\bibfield
  {journal} {\bibinfo  {journal} {arXiv}\ } (\bibinfo {year}
  {2020}{\natexlab{c}})},\ \Eprint {https://arxiv.org/abs/2001.11420}
  {arXiv:2001.11420 [astro-ph.CO]} \BibitemShut {NoStop}%
\bibitem [{\citenamefont {Nayak}\ and\ \citenamefont
  {Saha}(2021)}]{Nayak:2021lzf}%
  \BibitemOpen
  \bibfield  {author} {\bibinfo {author} {\bibfnamefont {P.}~\bibnamefont
  {Nayak}}\ and\ \bibinfo {author} {\bibfnamefont {R.}~\bibnamefont {Saha}},\
  }\bibfield  {title} {\bibinfo {title} {{Application of Genetic Algorithm to
  Estimate the Large Angular Scale Features of Cosmic Microwave Background}},\
  }\href@noop {} {\  (\bibinfo {year} {2021})},\ \Eprint
  {https://arxiv.org/abs/2102.06569} {arXiv:2102.06569 [astro-ph.CO]}
  \BibitemShut {NoStop}%
\end{thebibliography}%

\end{document}